\documentclass[aps,prb,superscriptaddress,amsmath,amssymb,reprint,floatfix]{revtex4-2}
\usepackage[colorlinks=true, urlcolor=blue, linkcolor=blue, citecolor=blue]{hyperref}
\usepackage[dvipsnames]{xcolor}
\usepackage[utf8]{inputenc}
\usepackage{graphicx}
\usepackage{bm}
\usepackage{booktabs}
\usepackage{natbib}

\newcommand{\s}{\bm{s}}

\newcommand{\g}{\mathbf{g}}

\begin{document}

\title{Universal sampling of spin systems across quenched disorder}

\author{Jing Liu}
\email{jing.liu@bupt.edu.cn}
\thanks{Equal contribution}
\affiliation{School of Physical Science and Technology, Beijing University of Posts and Telecommunications, Beijing 100876, China}
\affiliation{Institute of Theoretical Physics, Chinese Academy of Sciences, Beijing 100190, China}

\author{Yeyuan Wu}
\thanks{Equal contribution}
\affiliation{School of Fundamental Physics and Mathematical Sciences, Hangzhou Institute for Advanced Study, UCAS, Hangzhou 310024, China}
\affiliation{Institute of Theoretical Physics, Chinese Academy of Sciences, Beijing 100190, China}

\author{Ying Tang}
\email{jamestang23@gmail.com}
\affiliation{Institute of Fundamental and Frontier Sciences, University of Electronic Science and Technology of China, Chengdu 611731, China}
\affiliation{School of Physics, University of Electronic Science and Technology of China, Chengdu 611731, China}
\affiliation{Non-classical Information Science Basic Discipline Research Center of Sichuan Province, University of Electronic Science and Technology of China, Chengdu 611731, China}

\author{Pan Zhang}
\email{panzhang@itp.ac.cn}
\affiliation{Institute of Theoretical Physics, Chinese Academy of Sciences, Beijing 100190, China}
\affiliation{School of Fundamental Physics and Mathematical Sciences, Hangzhou Institute for Advanced Study, UCAS, Hangzhou 310024, China}
\affiliation{School of Physical Sciences, University of Chinese Academy of Sciences, Beijing 100049, China}

\begin{abstract}
Statistical physics extracts macroscopic laws by averaging over the many microscopic degrees of freedom of a system.
Disordered systems demand a second and far harder average, one over the quenched randomness itself.
The classic analytical routes, the replica and cavity methods, become uncontrolled outside mean-field or tree-like limits, and conventional numerical algorithms like parallel tempering require expensive, independent equilibration for every disorder realization.
In this work, we introduce a universal neural variational framework that amortizes inference across the disorder ensemble, eliminating both the need for per-instance Markov chain equilibration and the cost of retraining instance-specific variational ansatzes.
Built on an encoder-decoder Transformer architecture, after training once, it produces an explicit approximation to the Boltzmann distribution given previously unseen disorder realizations without further optimization.
We validate this framework on 2D Edwards-Anderson models, and apply it to the random-bond Ising model, successfully capturing the Binder cumulant crossings near the Nishimori multicritical point.
These results shift the object of variational inference from the single instance to the disorder ensemble, opening a route to frustrated many-body systems where instance-by-instance computation is prohibitive.
\end{abstract}

\maketitle

\paragraph*{Introduction---}
Evaluating quenched thermodynamic observables in disordered spin systems necessitates extensive averaging over numerous disorder realizations, posing a persistent computational challenge~\cite{sfedwards1975theory,binder1986spin}.
Parallel tempering Monte Carlo~\cite{hukushima1996exchange} remains the benchmark for sampling the rugged free-energy landscapes of spin glasses.
Although replica-exchange algorithms mitigate the critical slowing down of local updates, equilibration becomes prohibitively expensive deep within the low-temperature phase.
To circumvent these barriers, variational autoregressive networks (VANs)~\cite{wu2019solving} have been proposed for direct sampling, generating global proposals for Markov chains~\cite{mcnaughton2020boosting,nicoli2020asymptotically,wu2021unbiased,ciarella2023machinelearningassisted,delbono2026demonstrating}.
However, standard VANs exhibit a fundamental limitation when applied to quenched ensembles: training the neural network ansatz from scratch for each specific Hamiltonian negates their computational efficiency, rendering large-scale ensemble studies intractable~\cite{mcnaughton2020boosting,biazzo2024sparse,delbono2025nearestneighbors,delbono2026demonstrating,liu2025efficient,bialas2026sampling,zhong2026scalable}.

Generalizing inference across varying Hamiltonians requires universal meta-algorithms. 
While analytic frameworks like mean-field theory~\cite{kardar2007statistical}, Thouless-Anderson-Palmer equations~\cite{thouless1977solution}, and belief propagation~\cite{bethe1935statistical,yedidia2003understanding} yield instance-agnostic rules within strict validity domains, constructing deep generative models for zero-shot inference on arbitrary, highly frustrated discrete topologies remains inherently difficult. 
Machine learning addresses analogous bottlenecks via \textit{amortized inference}~\cite{kingma2014autoencoding,gershman2014amortized}, a paradigm that replaces iterative, instance-specific optimization with a direct functional mapping from system parameters to ensemble-optimized variational states.
This approach has shown early promise in adjacent domains: foundational neural-network quantum states (FNQS) generalize across Hamiltonians~\cite{rende2025foundation,viteritti2026quantum}, and transferable normalizing flows sample continuous molecular conformations~\cite{tan2025amortized,chen2026coarsegrained}.
For classical spin systems, VANs generalize across continuous temperatures to evaluate differentiable thermal observables~\cite{li2025deep}.
However, extending this amortized framework across distinct disorder realizations remains unsolved.
Unlike global temperature shifts, varying local disorder fundamentally restructures the discrete energy landscape, precluding zero-shot inference with standard generative models.

\begin{figure}[!t]
\centering
\includegraphics[width=\linewidth]{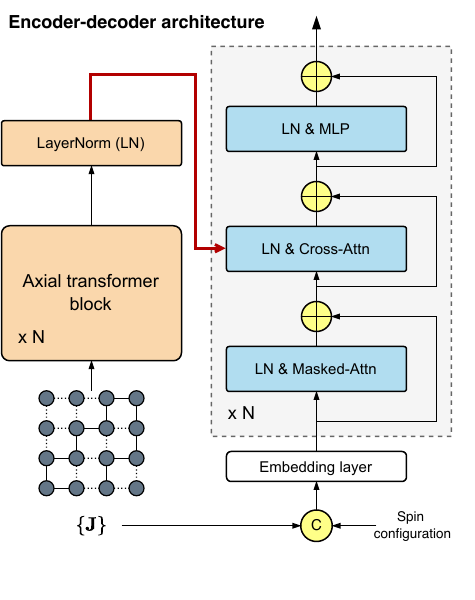}
\caption{Conceptual overview of the disorder-conditioned neural Boltzmann sampling framework. Left: The encoder maps the coupling matrix $\{\mathbf{J}\}$ to per-site embeddings $\{\mathbf{h}_i\}$ through $N$ axial transformer blocks, executing once per disorder realization. Right: The autoregressive decoder generates spin configurations patch by patch. Each patch token concatenates the spin embedding and local coupling features, then is updated via causal masked self-attention and cross-attention with $\{\mathbf{h}_i\}$ to parameterize the conditional distribution $q_\theta(\mathbf{s}|\mathbf{J})$.}
\label{fig:1}
\end{figure}

In this \textit{Letter}, we generalize the VAN framework from an instance-specific ansatz to a realization-independent meta-algorithm, avoiding per-instance retraining.
By processing Hamiltonian couplings to parameterize conditional spin distributions, our encoder-decoder architecture effectively models the equilibrium statistics of the full disorder ensemble.
Trained on dynamically generated instances, the model directly maps unseen disorder realizations to their Boltzmann distributions without retraining.
We test this zero-shot framework on 2D Edwards-Anderson models~\cite{sfedwards1975theory}, yielding accurate finite-temperature equilibrium sampling and ground-state energies under both bimodal and Gaussian disorder. 
We further apply this method to the random-bond Ising model~\cite{nishimori1981internal} to compute the quenched disorder-averaged Binder cumulant near the Nishimori multicritical point.

\paragraph*{Variational quenched-averaged free energy---}
In standard applications of VANs to disordered spin glass systems~\cite{mcnaughton2020boosting,biazzo2024sparse,delbono2025nearestneighbors,delbono2026demonstrating,liu2025efficient,bialas2026sampling}, equilibrium properties for a quenched disorder realization $J$ are obtained by introducing a parameterized approximation to the exact Boltzmann distribution $P_J(\s) = e^{-\beta E_J(\s)}/Z_J$.
Here, $E_J(\s) = -\sum_{\langle i,j \rangle} J_{ij} s_i s_j$ is the system Hamiltonian at inverse temperature $\beta=1/T$, and $Z_J = \sum_{\s} e^{-\beta E_J(\s)}$ denotes the associated partition function.
By employing an autoregressive neural network~\cite{larochelle2011neural,germain2015made,oord2016pixel,vaswani2017attention} $q_\theta(\s) = \prod_{i} q_\theta(s_i | \s_{<i})$ as the variational ansatz, one constructs an exactly normalized distribution over spin configurations $\s$, where $\s_{<i} = (s_1, \ldots, s_{i-1})$ denotes the causal history under a specified ordering.
This factorization permits direct ancestral sampling~\cite{bishop2006pattern}, circumventing the autocorrelation times inherent to Markov chain Monte Carlo methods~\cite{newman1999monte}.
The optimal parameters $\theta^*$ are obtained by minimizing the variational free energy, which strictly upper-bounds the exact free energy:
\begin{equation}
    \theta^* = \arg\min_\theta \mathbb{E}_{\s \sim q_\theta} \left[ E_J(\s) + \frac{1}{\beta} \ln q_\theta(\s) \right].
\end{equation}
Because $q_\theta(\s)$ takes only $\s$ as input, the disorder couplings bypass the forward pass and are instead implicitly encoded into the network parameters during optimization. 
Consequently, this instance-specific paradigm requires independent, computationally intensive training for each distinct disorder realization~\cite{liu2025efficient}.

To circumvent this bottleneck, we introduce an amortized inference scheme: a single conditional model $q_\theta(\s \vert J)$ maps an arbitrary realization $J \sim P(J)$ directly to its approximate Boltzmann distribution.
To train this mapping to generalize across different Hamiltonians, we optimize its parameters $\theta$ over the entire disorder ensemble $P(J)$. 
This naturally transforms the training objective into minimizing the variational quenched-averaged free energy:
\begin{equation}\label{eq:qvb}
    \mathcal{F}(\theta) = \mathbb{E}_{J \sim P(J)} \mathbb{E}_{\s \sim q_\theta(\cdot\vert J)} \left[ E_J(\s) + \frac{1}{\beta} \ln q_\theta(\s\vert J) \right].
\end{equation}
This bound rigorously satisfies $\mathcal{F}(\theta) \geq \mathbb{E}_{J \sim P(J)}[F_{\mathrm{true}}(J)]$, where $F_{\mathrm{true}}(J)=-(1/\beta) \ln Z_J$ is the exact free energy of a specific instance $J$. 
Consequently, minimizing $\mathcal{F}(\theta)$ tightens the variational free-energy estimate on average across the disorder ensemble. 
Once trained, the conditional ansatz evaluates variational free energies and generates independent samples for unseen realizations without retraining, effectively reducing computational costs.

\begin{figure*}[!t]
\centering
\includegraphics[width=\linewidth]{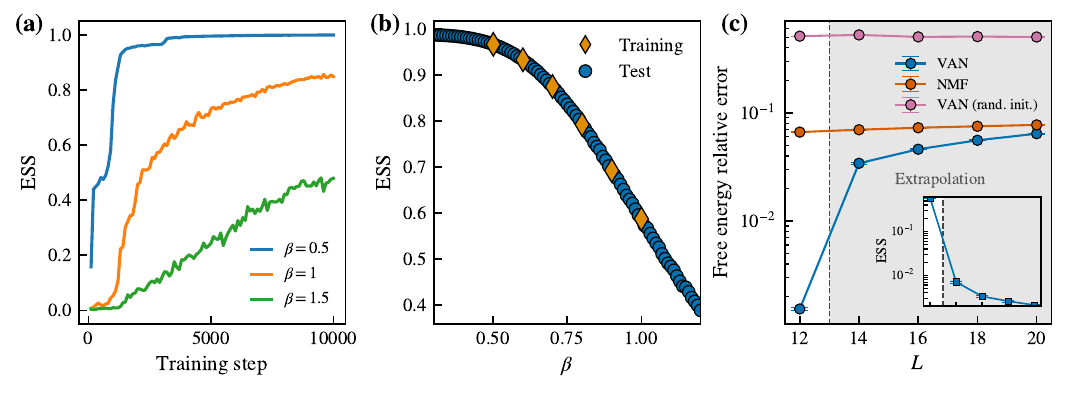}
\caption{Performance on the 2D Edwards-Anderson model with open boundary conditions ($L=12$).
(a) Training dynamics. The normalized effective sample size (ESS) is evaluated on a fixed validation set (20 realizations) during training for bimodal couplings.
(b) Thermal generalization. By conditioning the encoder on $K_{ij} = \beta J_{ij}$, a model trained on discrete $\beta$ values seamlessly interpolates and extrapolates to unseen continuous thermal regimes for Gaussian couplings without retraining.
(c) Extrapolation over system size. The model is trained on $L=12$ and evaluated on $L=\{12, 14, 16, 18, 20\}$ (gray shading marks the extrapolation regime $L>12$), compared with naive mean-field (NMF) and a randomly initialized model.}
\label{fig:2}
\end{figure*}

\paragraph*{Disorder-conditioned neural ansatz---}
While early VANs enforced the autoregressive property via masked dense or convolutional networks (e.g., NADE~\cite{larochelle2011neural}, MADE~\cite{germain2015made}, PixelCNN~\cite{oord2016pixel}), recent implementations predominantly utilize Transformers~\cite{liu2025efficient,bialas2026sampling,zhong2026scalable}. 
The Transformer architecture offers distinct computational advantages, including highly parallelizable training and superior scalability, while its attention mechanism naturally provides a global receptive field without imposing strict geometric priors~\cite{bialas2026sampling,zhong2026scalable}.
However, conventional Transformer VANs rely on a decoder-only architecture to parameterize $q_\theta(\s) = \prod_{i} q_\theta(s_i | \s_{<i})$, restricting the input space exclusively to physical spins.
To explicitly condition generation on quenched disorder, we introduce an encoder-decoder architecture~\cite{vaswani2017attention,fitzek2025rydberggpt}, see Fig.~\ref{fig:1}.
The encoder processes the coupling matrix $J$ via axial attention layers~\cite{ho2019axial}, which exploit the regular lattice geometry to efficiently establish a global receptive field.
This yields continuous per-site embeddings $\{\mathbf{h}_i\}_{i=1}^N$ that aggregate both local coupling topologies and global structural features of the disorder.
Because the encoder output is strictly $J$-dependent, the forward pass executes exactly once per disorder realization.

Because the Ising Hamiltonian exhibits local $\mathbb{Z}_2$ gauge invariance ($s_i \to \epsilon_i s_i$, $J_{ij} \to \epsilon_i \epsilon_j J_{ij}$ with $\epsilon_i \in \{-1, 1\}$)~\cite{wegner1971duality}, a single physical energy landscape corresponds to $2^{N-1}$ formally distinct coupling matrices.
To facilitate efficient representation learning, deterministic gauge-fixing is applied to collapse this redundancy into a canonical coupling matrix $J$~\cite{fan2023searching}.

Autoregressive sampling then proceeds via a causal Transformer decoder. 
To efficiently capture local correlations and reduce the effective sequence length, we coarse-grain spins into patches (typically $2\times 2$)~\cite{dosovitskiy2021image}. 
For the $t$-th patch, the decoder state $\g_t \in \mathbb{R}^{d_{\mathrm{emb}}}$, where $d_{\mathrm{emb}}$ represents the embedding dimension, is initialized by concatenating the spin token embedding $\mathbf{e}_s^{(t)}$ and a local coupling feature embedding $\mathbf{e}_J^{(t)}$: $\g_t = \operatorname{Concat}(\mathbf{e}_s^{(t)}, \mathbf{e}_J^{(t)})$.
This concatenation is also employed in FNQS~\cite{rende2025foundation,viteritti2026quantum}, and our ablations confirm that this explicit concatenation systematically outperforms standard token embedding alone.
The decoder updates its hidden state $\g_t$ via masked self-attention over the causal history and cross-attention with the disorder embeddings $\{\mathbf{h}_i\}$.
Formally, these attention mechanisms compute the weight for each position pair as
\begin{equation}
    \alpha_{tj} = \frac{\exp(\mathbf{q}_t \cdot \mathbf{k}_j / \sqrt{d_k})}{\sum_i \exp(\mathbf{q}_t \cdot \mathbf{k}_i / \sqrt{d_k})},
\end{equation}
yielding $\operatorname{Attn}(\mathbf{q}_t, \mathbf{K}, \mathbf{V}) = \sum_j \alpha_{tj} \mathbf{v}_j$, with projection dimension $d_k$.
For masked self-attention, queries, keys, and values are linearly projected from the decoder states ($\mathbf{q}_t = W^Q \g_t$, $\mathbf{k}_j = W^K \g_j$, $\mathbf{v}_j = W^V \g_j$).
Strict causality is enforced by restricting the softmax sum to positions $\le t$ and setting $\alpha_{tj}=0$ for $j>t$.
The positional information is incorporated through rotary position embedding (RoPE)~\cite{su2024roformer} directly in the attention calculation.
In cross-attention, queries are derived from the decoder ($\mathbf{q}_t = W^Q_c \g_t$), whereas keys and values are projected from the encoder outputs ($\mathbf{k}_j = W^K_c \mathbf{h}_j$, $\mathbf{v}_j = W^V_c \mathbf{h}_j$), with the softmax normalization taken over all $N$ positions.
This cross-attention mechanism explicitly couples the learned disorder environment to the autoregressive generation process.

The encoder and decoder are jointly optimized to minimize $\mathcal{F}(\theta)$ using the REINFORCE gradient estimator~\cite{williams1992simple}. 
To ensure robust generalization across the disorder ensemble $P(J)$, new realizations are independently sampled at each training iteration. 
Gauge-fixing details, architectural specifications, and surrogate loss derivations are provided in the Supplemental Material~\cite{sm}.

\paragraph*{Zero-shot inference on the Edwards-Anderson model---}
We first demonstrate our realization-independent framework on the 2D EA model, governed by the Hamiltonian $E_J(\s) = - \sum_{\langle i,j \rangle} J_{ij} s_i s_j$, where nearest-neighbor couplings $J_{ij}$ are independently drawn from bimodal ($P(J_{ij}) = \frac{1}{2}[\delta(J_{ij}-1) + \delta(J_{ij}+1)]$) or Gaussian ($P(J_{ij}) = \frac{1}{\sqrt{2\pi}} \exp(-J_{ij}^2/2)$) distributions. 
We employ open boundary conditions (OBC), permitting exact calculation of the instance-specific free energy~\cite{kac1952combinatorial}. 
Additionally, sampling efficiency can be measured via the normalized effective sample size (ESS)~\cite{elvira2022rethinking} $\text{ESS} = \langle w_i \rangle^2 / \langle w_i^2 \rangle$, where $w_i = \exp(-\beta E_J(\s_i)) / q_\theta(\s_i | J)$ denotes the importance weight of configuration $\s_i$. 
By definition, $0 < \text{ESS} \le 1$, and $\text{ESS}=1$ if $q_\theta(\s|J)$ exactly recovers the true Boltzmann distribution $P_J(\s)$. 
To precisely monitor the training dynamics, we periodically evaluate the model on a fixed validation set.
As training progresses, a systematic increase in the ESS is observed [Fig.~\ref{fig:2}(a)].
The training dynamics also exhibit a clear temperature dependence: convergence becomes noticeably slower and the asymptotic ESS becomes lower as the temperature decreases. 
This behavior reflects the increasingly rugged free-energy landscape that systematically impedes optimization.

We then investigate how well the model generalizes across different temperatures.
This is done by conditioning the encoder on the dimensionless local coupling $K_{ij} = \beta J_{ij}$.
Figure~\ref{fig:2}(b) demonstrates that a model trained exclusively on a sparse, discrete set of $\beta$ values can interpolate and extrapolate to unseen continuous thermal regimes for Gaussian couplings. 
The model maintains high sampling efficiency across the temperature range, successfully capturing the relevant statistical properties without requiring separate, temperature-specific models.
This thermal generalization extends beyond the variational temperature-differentiable (VaTD) framework in Ref.~\cite{li2025deep}, which conditions a generative model on the global scalar $\beta$ for a fixed Hamiltonian.
Our encoder-decoder architecture reduces to the VaTD setting when the disorder distribution collapses to a uniform coupling, making the latter the disorder-free special case of our framework.

We next examine the framework's capability to extrapolate to larger system sizes.
Because the network's internal operations, including the encoder's feature extraction and the decoder's autoregressive generation, are independent of the global lattice dimension, our architecture is inherently size-agnostic. 
Figure~\ref{fig:2}(c) illustrates this extrapolation over lattice size at $\beta=1.0$ for Gaussian disorder.
A model trained on $L=12$ is directly evaluated on unseen larger lattices.
We observe that the relative error remains lower than that of the naive mean-field (NMF) approximation.
The ability to outperform the NMF baseline indicates that the network successfully extracts and transfers non-local spin correlations across different length scales.

\begin{figure}[!t]
\centering
\includegraphics[width=\linewidth]{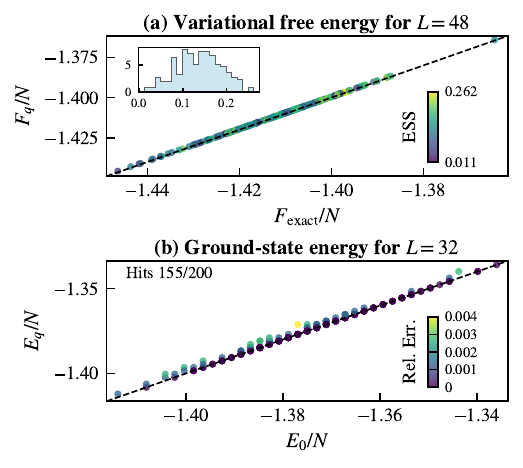}
\caption{Results obtained via curriculum training, where the model is trained sequentially on system sizes $L \in \{8, 12, 16, 24, 32, 40, 48\}$, each stage warm-started from the converged weights of the previous size.
(a) Variational free energies benchmarked against exact Kac-Ward solutions across an independent test set (200 realizations) for Gaussian disorder at $L=48$, $\beta=1.0$.
(b) Ground-state search for bimodal couplings at $L=32$ via variational annealing from the curriculum-trained model, showing the minimum energy found versus the exact ground-state energy obtained by \cite{pardella2008exact}.}
\label{fig:3}
\end{figure}

\begin{figure}[!t]
\centering
\includegraphics[width=\linewidth]{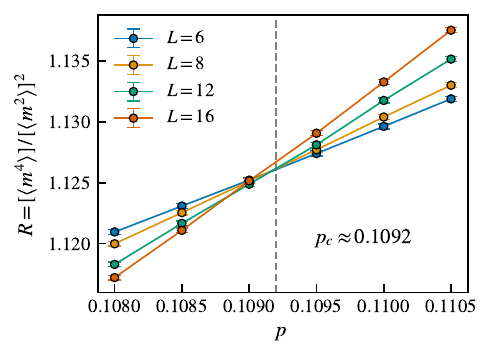}
\caption{Quenched-averaged Binder cumulant $U = [\langle m^4 \rangle] / [\langle m^2 \rangle]^2$ in the 2D random-bond Ising model as a function of the antiferromagnetic bond probability $p$ for various system sizes $L$.
Periodic boundary conditions are adopted to suppress finite-size effects.
The curves cross near the expected multicritical Nishimori point $p \approx 0.1092$.
For each point, observables are computed via importance sampling across $10^6$ distinct disorder realizations using our amortized model.}
\label{fig:4}
\end{figure}

\paragraph*{Curriculum training and ground-state search---}
Although the direct zero-shot transfer degrades at larger $L$, the fact that a model trained on a smaller lattice still outperforms the NMF baseline on larger ones indicates that the learned representations capture non-local correlations transferable across length scales.
This suggests that a model pre-trained at a smaller size provides a markedly better initialization than random weights for larger systems [Fig.~\ref{fig:2}(c)].
Exploiting this, we adopt a curriculum learning strategy~\cite{bengio2009curriculum}: starting from a converged model at a small lattice size, we retrain at progressively larger $L$, each time warm-starting from the model obtained at the previous size, a strategy also employed in related fields~\cite{roth2020iterative,casert2021dynamical,merali2026parallel}.
As shown in Fig.~\ref{fig:3}(a), this approach achieves highly accurate free-energy estimations on systems up to $L=48$ at $\beta=1.0$.
This represents a substantial scale, especially considering that conventional VANs restricted to single-instance optimization have recently reached $L=64$~\cite{zhong2026scalable,bialas2026sampling}.
We further extend this framework to search for ground states by introducing an annealing schedule~\cite{hibat-allah2021variational,delbono2026demonstrating,zhong2026scalable} that forces the network, during training, to concentrate its probability mass on the lowest-energy configurations. 
As shown in Fig.~\ref{fig:3}(b), this annealed network rapidly generates high-quality ground-state candidates for novel instances, offering a highly efficient approach to low-energy sampling in rugged landscapes.

\paragraph*{Multicritical Nishimori point in the random-bond Ising model---}
Finally, we demonstrate the utility of our method by investigating the critical behavior of the 2D random-bond Ising model~\cite{nishimori1981internal}.
The system is characterized by the disorder distribution $P(J_{ij}) = (1-p)\delta(J_{ij}-1) + p\delta(J_{ij}+1)$, where $p$ represents the probability of antiferromagnetic bonds.
Simulations are performed on the Nishimori line~\cite{nishimori1981internal}, where the inverse temperature $\beta$ is fixed by $p$ via $e^{-2\beta} = p / (1-p)$.
This model features a multicritical Nishimori point (MNP) at $p \approx 0.1092$~\cite{hasenbusch2008multicritical,parisentoldin2009strongdisorder,chen2025tensor}, where dimensionless scaling quantities, such as the Binder cumulant, become scale-invariant and curves for different system sizes intersect at a common value. 
Reliable quenched averages of the Binder cumulant demand $\sim 10^6$ disorder realizations per parameter point.
Parallel tempering requires independent equilibration for each realization, making an ensemble study of this scale expensive.
Standard VANs inherit the same bottleneck because the model must be retrained independently for each realization~\cite{zhong2026scalable}.
Our amortized model, by contrast, is trained once per $p$ and then generates i.i.d.~samples across an extensive ensemble of distinct instances, eliminating both the equilibration overhead of MCMC and the retraining cost of instance-specific VANs. 
The quenched-averaged observables $[\langle m^2 \rangle]^2$ and $[\langle m^4 \rangle]$ are then computed via importance sampling~\cite{nicoli2020asymptotically}. 
The resulting Binder cumulant $U = [\langle m^4 \rangle] / [\langle m^2 \rangle]^2$ (Fig.~\ref{fig:4}) exhibits a clear crossing consistent with the known MNP. 
This result not only successfully captures a highly non-trivial critical phenomenon but also underscores the broader potential to solve challenging problems in complex disordered systems.
For details of the training and more extended numerical results, see Supplemental Material~\cite{sm}.

\paragraph*{Discussion and Conclusion---}
In summary, we have introduced an amortized variational framework that generalizes neural Boltzmann sampling across quenched disorder realizations. 
The encoder-decoder architecture learns a direct mapping from Hamiltonian couplings to equilibrium statistics, enabling zero-shot inference on unseen disorder instances without retraining. 
We validated this approach on the 2D EA model, achieving accurate free energies up to $L=48$ and ground-state energies up to $L=32$ via curriculum training. 
We further demonstrated its utility by computing the quenched-averaged Binder cumulant in the random-bond Ising model, successfully resolving the multicritical Nishimori point from $10^6$ disorder realizations.

Looking forward, this amortized framework can be extended in several directions.
On the model side, the sampling cost can be further reduced by replacing the vanilla attention mechanism with linear-complexity alternatives~\cite{peng2023rwkv,katharopoulos2020transformers} or complementing hierarchical sampling schemes~\cite{bialas2022hierarchical,singha2025multilevel,singha2026scalable,bialas2026hierarchical,du2026scaling} to enable access to substantially larger systems.
On the application side, vanilla VANs have been used to model the stochastic dynamics~\cite{tang2023neuralnetwork,gao2024generative,tang2024learning,zhao2025nonequilibrium,weng2025tracking,liu2026characterizing}, and conditioning such models on time-dependent or model-dependent evolution operators would enable zero-shot prediction of time-evolved classical distributions.
A similar operator-learning approach has recently been demonstrated for driven quantum dynamics~\cite{qi2026neural}.
For neural-network quantum states, while the current state-of-the-art is based on vision transformers~\cite{rende2025foundation,viteritti2026quantum}, our autoregressive approach offers the distinct advantage of exactly normalized wave-function amplitudes, making the extension of this amortized framework to quantum spin glasses a natural next step.

\paragraph*{Acknowledgments---}
This work is supported by the National Natural Science Foundation of China [Grant No.: 12405047, 12325501, 12322501, 12575035], the Strategic Priority Research Program of Chinese Academy of Sciences [Grant No.XDB1680000], and the Fundamental Research Funds for the Central Universities.

\paragraph*{Data availability---}
The source code will be openly available upon acceptance.

\bibliography{references}

@inproceedings{bengio2009curriculum,
  title = {{Curriculum learning}},
  author = {Bengio, Yoshua and Louradour, J{\'e}r{\^o}me and Collobert, Ronan and Weston, Jason},
  year = {2009},
  booktitle = {Proceedings of the 26th Annual International Conference on Machine Learning, ICML 2009, Montreal, Quebec, Canada, June 14-18, 2009},
  publisher = {ACM},
  series = {ACM International Conference Proceeding Series},
  volume = {382},
  pages = {41--48},
  doi = {10.1145/1553374.1553380},
  url = {https://doi.org/10.1145/1553374.1553380},
  editor = {Danyluk, Andrea Pohoreckyj and Bottou, L{\'e}on and Littman, Michael L.}
}

@article{bethe1935statistical,
  title = {{Statistical theory of superlattices}},
  author = {Bethe, Hans A},
  year = {1935},
  journal = {Proceedings of the Royal Society of London. Series A-Mathematical and Physical Sciences},
  publisher = {The Royal Society London},
  volume = {150},
  number = {871},
  pages = {552--575}
}

@article{bialas2022hierarchical,
  title = {{Hierarchical Autoregressive Neural Networks for Statistical Systems}},
  author = {Bia{\l}as, Piotr and Korcyl, Piotr and Stebel, Tomasz},
  year = {2022},
  journal = {Computer Physics Communications},
  volume = {281},
  pages = {108502},
  doi = {10.1016/j.cpc.2022.108502},
  issn = {0010-4655},
  url = {https://www.sciencedirect.com/science/article/pii/S0010465522002211}
}

@article{bialas2026hierarchical,
  title = {{Hierarchical Autoregressive Neural Networks in Three-Dimensional Statistical System}},
  author = {Bia{\l}as, Piotr and Chahar, Vaibhav and Korcyl, Piotr and Stebel, Tomasz and Winiarski, Mateusz and Zapolski, Dawid},
  year = {2026},
  journal = {Computer Physics Communications},
  volume = {318},
  pages = {109892},
  doi = {10.1016/j.cpc.2025.109892},
  issn = {0010-4655},
  url = {https://www.sciencedirect.com/science/article/pii/S0010465525003947}
}

@misc{bialas2026sampling,
  title = {{Sampling two-dimensional spin systems with transformers}},
  author = {Bia{\l}as, Piotr and Korcyl, Piotr and Stebel, Tomasz and Stefa{\'n}ski, Adam and Zapolski, Dawid},
  year = {2026},
  url = {https://arxiv.org/abs/2604.27738},
  eprint = {2604.27738},
  archiveprefix = {arXiv},
  primaryclass = {cond-mat.dis-nn}
}

@article{biazzo2024sparse,
  title = {{Sparse autoregressive neural networks for classical spin systems}},
  author = {Biazzo, Indaco and Wu, Dian and Carleo, Giuseppe},
  year = {2024},
  journal = {Machine Learning: Science and Technology},
  volume = {5},
  number = {2},
  pages = {025074},
  doi = {10.1088/2632-2153/ad5783},
  issn = {2632-2153},
  url = {https://iopscience.iop.org/article/10.1088/2632-2153/ad5783},
  urldate = {2024-08-02},
  langid = {english}
}

@article{binder1986spin,
  title = {{Spin glasses: Experimental facts, theoretical concepts, and open questions}},
  author = {Binder, K. and Young, A. P.},
  year = {1986},
  journal = {Reviews of Modern Physics},
  publisher = {American Physical Society},
  volume = {58},
  number = {4},
  pages = {801--976},
  doi = {10.1103/RevModPhys.58.801},
  url = {https://link.aps.org/doi/10.1103/RevModPhys.58.801}
}

@book{bishop2006pattern,
  title = {{Pattern Recognition and Machine Learning}},
  author = {Bishop, Christopher M},
  year = {2006},
  publisher = {Springer New York, NY}
}

@article{casert2021dynamical,
  title = {{Dynamical Large Deviations of Two-Dimensional Kinetically Constrained Models Using a Neural-Network State Ansatz}},
  author = {Casert, Corneel and Vieijra, Tom and Whitelam, Stephen and Tamblyn, Isaac},
  year = {2021},
  journal = {Physical Review Letters},
  publisher = {American Physical Society},
  volume = {127},
  number = {12},
  pages = {120602},
  doi = {10.1103/PhysRevLett.127.120602},
  url = {https://link.aps.org/doi/10.1103/PhysRevLett.127.120602}
}

@article{chen2025tensor,
  title = {{Tensor network Monte Carlo simulations for the two-dimensional random-bond Ising model}},
  author = {Chen, Tao and Guo, Erdong and Zhang, Wanzhou and Zhang, Pan and Deng, Youjin},
  year = {2025},
  journal = {Physical Review B},
  publisher = {American Physical Society},
  volume = {111},
  number = {9},
  pages = {094201},
  doi = {10.1103/PhysRevB.111.094201},
  url = {https://link.aps.org/doi/10.1103/PhysRevB.111.094201}
}

@misc{chen2026coarsegrained,
  title = {{Coarse-Grained Boltzmann Generators}},
  author = {Chen, Weilong and Zhao, Bojun and Eckwert, Jan and Zavadlav, Julija},
  year = {2026},
  url = {https://arxiv.org/abs/2602.10637},
  eprint = {2602.10637},
  archiveprefix = {arXiv},
  primaryclass = {cs.LG}
}

@article{ciarella2023machinelearningassisted,
  title = {{Machine-learning-assisted Monte Carlo fails at sampling computationally hard problems}},
  author = {Ciarella, Simone and Trinquier, Jeanne and Weigt, Martin and Zamponi, Francesco},
  year = {2023},
  journal = {Machine Learning: Science and Technology},
  volume = {4},
  number = {1},
  pages = {010501},
  doi = {10.1088/2632-2153/acbe91},
  issn = {2632-2153},
  url = {https://iopscience.iop.org/article/10.1088/2632-2153/acbe91},
  urldate = {2023-04-10},
  langid = {english}
}

@article{delbono2025nearestneighbors,
  title = {{Nearest-neighbors neural network architecture for efficient sampling of statistical physics models}},
  author = {Del Bono, Luca Maria and {Ricci-Tersenghi}, Federico and Zamponi, Francesco},
  year = {2025},
  journal = {Machine Learning: Science and Technology},
  publisher = {IOP Publishing},
  volume = {6},
  number = {2},
  pages = {025029},
  doi = {10.1088/2632-2153/adcdc1},
  issn = {2632-2153},
  url = {https://doi.org/10.1088/2632-2153/adcdc1}
}

@article{delbono2026demonstrating,
  title = {{Demonstrating real advantage of machine learning--enhanced Monte Carlo for combinatorial optimization}},
  author = {Del Bono, Luca Maria and {Ricci-Tersenghi}, Federico and Zamponi, Francesco},
  year = {2026},
  journal = {Proceedings of the National Academy of Sciences},
  publisher = {Proceedings of the National Academy of Sciences},
  volume = {123},
  number = {19},
  pages = {e2534768123},
  doi = {10.1073/pnas.2534768123},
  url = {https://doi.org/10.1073/pnas.2534768123},
  urldate = {2026-06-09}
}

@inproceedings{dosovitskiy2021image,
  title = {{An Image is Worth 16x16 Words: Transformers for Image Recognition at Scale}},
  author = {Dosovitskiy, Alexey and Beyer, Lucas and Kolesnikov, Alexander and Weissenborn, Dirk and Zhai, Xiaohua and Unterthiner, Thomas and Dehghani, Mostafa and Minderer, Matthias and Heigold, Georg and Gelly, Sylvain and Uszkoreit, Jakob and Houlsby, Neil},
  year = {2021},
  booktitle = {9th International Conference on Learning Representations, ICLR 2021, Virtual Event, Austria, May 3-7, 2021},
  publisher = {OpenReview.net},
  url = {https://openreview.net/forum?id=YicbFdNTTy},
  bibsource = {dblp computer science bibliography, https://dblp.org}
}

@misc{du2026scaling,
  title = {{Scaling Autoregressive Models for Lattice Thermodynamics}},
  author = {Du, Xiaochen and Nam, Juno and Liu, Sulin and {G{\'o}mez-Bombarelli}, Rafael},
  year = {2026},
  url = {https://arxiv.org/abs/2603.14695},
  eprint = {2603.14695},
  primaryclass = {cond-mat.stat-mech},
  archiveprefix = {arXiv}
}

@article{elvira2022rethinking,
  title = {{Rethinking the Effective Sample Size}},
  author = {Elvira, V{\'i}ctor and Martino, Luca and Robert, Christian P.},
  year = {2022},
  journal = {International Statistical Review},
  publisher = {John Wiley \& Sons, Ltd},
  volume = {90},
  number = {3},
  pages = {525--550},
  doi = {10.1111/insr.12500},
  issn = {0306-7734},
  url = {https://doi.org/10.1111/insr.12500},
  urldate = {2026-06-15}
}

@article{fan2023searching,
  title = {{Searching for spin glass ground states through deep reinforcement learning}},
  author = {Fan, Changjun and Shen, Mutian and Nussinov, Zohar and Liu, Zhong and Sun, Yizhou and Liu, Yang-Yu},
  year = {2023},
  journal = {Nature Communications},
  volume = {14},
  number = {1},
  pages = {725},
  doi = {10.1038/s41467-023-36363-w},
  issn = {2041-1723},
  url = {https://www.nature.com/articles/s41467-023-36363-w},
  urldate = {2023-02-19},
  langid = {english}
}

@article{fitzek2025rydberggpt,
  title = {{RydbergGPT}},
  author = {Fitzek, David and Hong Teoh, Yi and Cyrus Fung, H P and Dagnew, Gebremedhin A and Merali, Ejaaz and Schuyler Moss, M and MacLellan, Benjamin and Melko, Roger G},
  year = {2025},
  journal = {Machine Learning: Science and Technology},
  publisher = {IOP Publishing},
  volume = {6},
  number = {4},
  pages = {045057},
  doi = {10.1088/2632-2153/ae1d0b},
  issn = {2632-2153},
  url = {https://doi.org/10.1088/2632-2153/ae1d0b}
}

@article{gao2024generative,
  title = {{Generative Learning for Forecasting the Dynamics of High-Dimensional Complex Systems}},
  author = {Gao, Han and Kaltenbach, Sebastian and Koumoutsakos, Petros},
  year = {2024},
  journal = {Nature Communications},
  volume = {15},
  number = {1},
  pages = {8904},
  doi = {10.1038/s41467-024-53165-w},
  issn = {2041-1723},
  url = {https://doi.org/10.1038/s41467-024-53165-w}
}

@inproceedings{germain2015made,
  title = {{MADE: Masked Autoencoder for Distribution Estimation}},
  author = {Germain, Mathieu and Gregor, Karol and Murray, Iain and Larochelle, Hugo},
  year = {2015},
  booktitle = {Proceedings of the 32nd International Conference on Machine Learning},
  publisher = {PMLR},
  address = {Lille, France},
  series = {Proceedings of Machine Learning Research},
  volume = {37},
  pages = {881--889},
  url = {https://proceedings.mlr.press/v37/germain15.html},
  editor = {Bach, Francis and Blei, David}
}

@inproceedings{gershman2014amortized,
  title = {{Amortized inference in probabilistic reasoning}},
  author = {Gershman, Samuel and Goodman, Noah},
  year = {2014},
  booktitle = {Proceedings of the annual meeting of the cognitive science society},
  volume = {36},
  url = {https://escholarship.org/uc/item/34j1h7k5}
}

@article{hasenbusch2008multicritical,
  title = {{Multicritical Nishimori point in the phase diagram of the $\pm {J}$ Ising model on a square lattice}},
  author = {Hasenbusch, Martin and Toldin, Francesco Parisen and Pelissetto, Andrea and Vicari, Ettore},
  year = {2008},
  journal = {Physical Review E},
  publisher = {American Physical Society},
  volume = {77},
  number = {5},
  pages = {051115},
  doi = {10.1103/PhysRevE.77.051115},
  url = {https://link.aps.org/doi/10.1103/PhysRevE.77.051115}
}

@article{hibat-allah2021variational,
  title = {{Variational neural annealing}},
  author = {{Hibat-Allah}, Mohamed and Inack, Estelle M. and Wiersema, Roeland and Melko, Roger G. and Carrasquilla, Juan},
  year = {2021},
  journal = {Nature Machine Intelligence},
  volume = {3},
  number = {11},
  pages = {952--961},
  doi = {10.1038/s42256-021-00401-3},
  issn = {2522-5839},
  url = {https://www.nature.com/articles/s42256-021-00401-3},
  urldate = {2022-05-30},
  langid = {english}
}

@misc{ho2019axial,
  title = {{Axial Attention in Multidimensional Transformers}},
  author = {Ho, Jonathan and Kalchbrenner, Nal and Weissenborn, Dirk and Salimans, Tim},
  year = {2019},
  url = {https://arxiv.org/abs/1912.12180},
  eprint = {1912.12180},
  archiveprefix = {arXiv},
  primaryclass = {cs.CV}
}

@article{hukushima1996exchange,
  title = {{Exchange Monte Carlo Method and  Application to Spin Glass Simulations}},
  author = {Hukushima, Koji and Nemoto, Koji},
  year = {1996},
  journal = {Journal of the Physical Society of Japan},
  publisher = {The Physical Society of Japan},
  volume = {65},
  number = {6},
  pages = {1604--1608},
  doi = {10.1143/JPSJ.65.1604},
  issn = {0031-9015},
  url = {https://doi.org/10.1143/JPSJ.65.1604},
  urldate = {2025-06-24}
}

@article{kac1952combinatorial,
  title = {{A Combinatorial Solution of the Two-Dimensional Ising Model}},
  author = {Kac, M. and Ward, J. C.},
  year = {1952},
  journal = {Physical Review},
  publisher = {American Physical Society},
  volume = {88},
  number = {6},
  pages = {1332--1337},
  doi = {10.1103/PhysRev.88.1332},
  url = {https://link.aps.org/doi/10.1103/PhysRev.88.1332}
}

@book{kardar2007statistical,
  title = {{Statistical physics of fields}},
  author = {Kardar, Mehran},
  year = {2007},
  publisher = {Cambridge University Press}
}

@inproceedings{katharopoulos2020transformers,
  title = {{Transformers Are RNNs: Fast Autoregressive Transformers with Linear Attention}},
  author = {Katharopoulos, Angelos and Vyas, Apoorv and Pappas, Nikolaos and Fleuret, Fran{\c c}ois},
  year = {2020},
  booktitle = {Proceedings of the 37th International Conference on Machine Learning},
  publisher = {PMLR},
  series = {Proceedings of Machine Learning Research},
  volume = {119},
  pages = {5156--5165},
  url = {https://proceedings.mlr.press/v119/katharopoulos20a.html},
  editor = {III, Hal Daum{\'e} and Singh, Aarti}
}

@inproceedings{kingma2014autoencoding,
  title = {{Auto-Encoding Variational Bayes}},
  author = {Kingma, Diederik P. and Welling, Max},
  year = {2014},
  booktitle = {2nd International Conference on Learning Representations, ICLR 2014, Banff, AB, Canada, April 14-16, 2014, Conference Track Proceedings},
  url = {http://arxiv.org/abs/1312.6114},
  editor = {Bengio, Yoshua and LeCun, Yann}
}

@inproceedings{larochelle2011neural,
  title = {{The Neural Autoregressive Distribution Estimator}},
  author = {Larochelle, Hugo and Murray, Iain},
  year = {2011},
  booktitle = {Proceedings of the Fourteenth International Conference on Artificial Intelligence and Statistics},
  publisher = {PMLR},
  address = {Fort Lauderdale, FL, USA},
  series = {Proceedings of Machine Learning Research},
  volume = {15},
  pages = {29--37},
  url = {https://proceedings.mlr.press/v15/larochelle11a.html},
  editor = {Gordon, Geoffrey and Dunson, David and Dud{\'i}k, Miroslav}
}

@article{li2025deep,
  title = {{Deep Generative Modeling of the Canonical Ensemble with Differentiable Thermal Properties}},
  author = {Li, Shuo-Hui and Zhang, Yao-Wen and Pan, Ding},
  year = {2025},
  journal = {Physical Review Letters},
  publisher = {American Physical Society},
  volume = {135},
  number = {2},
  pages = {027301},
  doi = {10.1103/8wx7-kyx8},
  url = {https://link.aps.org/doi/10.1103/8wx7-kyx8}
}

@article{liu2025efficient,
  title = {{Efficient optimization of variational autoregressive networks with natural gradient}},
  author = {Liu, Jing and Tang, Ying and Zhang, Pan},
  year = {2025},
  journal = {Physical Review E},
  publisher = {American Physical Society},
  volume = {111},
  number = {2},
  pages = {025304},
  doi = {10.1103/PhysRevE.111.025304},
  url = {https://link.aps.org/doi/10.1103/PhysRevE.111.025304}
}

@misc{liu2026characterizing,
  title = {{Characterizing Full Nonequilibrium Dynamics of Simple Exclusion Processes}},
  author = {Zhimao Liu and Jing Liu and Pan Zhang and Ying Tang},
  year = {2026},
  url = {https://arxiv.org/abs/2608.25606},
  eprint = {2608.25606},
  archiveprefix = {arXiv},
  primaryclass = {cond-mat.stat-mech}
}

@article{mcnaughton2020boosting,
  title = {{Boosting Monte Carlo simulations of spin glasses using autoregressive neural networks}},
  author = {McNaughton, B. and Milo{\v s}evi{\'c}, M. V. and Perali, A. and Pilati, S.},
  year = {2020},
  journal = {Physical Review E},
  volume = {101},
  number = {5},
  pages = {053312},
  doi = {10.1103/PhysRevE.101.053312},
  issn = {2470-0045, 2470-0053},
  url = {https://link.aps.org/doi/10.1103/PhysRevE.101.053312},
  urldate = {2022-05-22},
  langid = {english}
}

@misc{merali2026parallel,
  title = {{Parallel Scan Recurrent Neural Quantum States for Scalable Variational Monte Carlo}},
  author = {Merali, Ejaaz and {Hibat-Allah}, Mohamed and Kohandel, Mohammad and Scalettar, Richard T. and Khatami, Ehsan},
  year = {2026},
  url = {https://arxiv.org/abs/2605.13807},
  eprint = {2605.13807},
  archiveprefix = {arXiv},
  primaryclass = {cond-mat.str-el}
}

@book{newman1999monte,
  title = {{Monte Carlo Methods in Statistical Physics}},
  author = {Newman, M.E.J. and Barkema, G.T.},
  year = {1999},
  publisher = {Clarendon Press},
  isbn = {978-0-19-851796-2},
  url = {https://books.google.co.jp/books?id=KKL2nQEACAAJ},
  lccn = {99213405}
}

@article{nicoli2020asymptotically,
  title = {{Asymptotically unbiased estimation of physical observables with neural samplers}},
  author = {Nicoli, Kim A. and Nakajima, Shinichi and Strodthoff, Nils and Samek, Wojciech and M{\"u}ller, Klaus-Robert and Kessel, Pan},
  year = {2020},
  journal = {Physical Review E},
  volume = {101},
  number = {2},
  pages = {023304},
  doi = {10.1103/PhysRevE.101.023304},
  issn = {2470-0045, 2470-0053},
  url = {https://link.aps.org/doi/10.1103/PhysRevE.101.023304},
  urldate = {2022-01-03},
  langid = {english}
}

@article{nishimori1981internal,
  title = {{Internal Energy, Specific Heat and Correlation Function of the Bond-Random Ising Model}},
  author = {Nishimori, Hidetoshi},
  year = {1981},
  journal = {Progress of Theoretical Physics},
  volume = {66},
  number = {4},
  pages = {1169--1181},
  doi = {10.1143/PTP.66.1169},
  issn = {0033-068X},
  url = {https://doi.org/10.1143/PTP.66.1169},
  urldate = {2025-06-25}
}

@inproceedings{oord2016pixel,
  title = {{Pixel Recurrent Neural Networks}},
  author = {van den Oord, A{\"a}ron and Kalchbrenner, Nal and Kavukcuoglu, Koray},
  year = {2016},
  booktitle = {Proceedings of the 33nd International Conference on Machine Learning, ICML 2016, New York City, NY, USA, June 19-24, 2016},
  publisher = {JMLR.org},
  series = {JMLR Workshop and Conference Proceedings},
  volume = {48},
  pages = {1747--1756},
  url = {http://proceedings.mlr.press/v48/oord16.html},
  editor = {Balcan, Maria-Florina and Weinberger, Kilian Q.}
}

@article{pardella2008exact,
  title = {{Exact Ground States of Large Two-Dimensional Planar Ising Spin Glasses}},
  author = {Pardella, G. and Liers, F.},
  year = {2008},
  journal = {Physical Review E},
  publisher = {American Physical Society},
  volume = {78},
  number = {5},
  pages = {056705},
  doi = {10.1103/PhysRevE.78.056705},
  url = {https://link.aps.org/doi/10.1103/PhysRevE.78.056705}
}

@article{parisentoldin2009strongdisorder,
  title = {{Strong-Disorder Paramagnetic-Ferromagnetic Fixed Point in the Square-Lattice $\pm {J}$ Ising Model}},
  author = {Parisen Toldin, Francesco and Pelissetto, Andrea and Vicari, Ettore},
  year = {2009},
  journal = {Journal of Statistical Physics},
  volume = {135},
  number = {5},
  pages = {1039--1061},
  doi = {10.1007/s10955-009-9705-5},
  issn = {1572-9613},
  url = {https://doi.org/10.1007/s10955-009-9705-5}
}

@inproceedings{peng2023rwkv,
  title = {{RWKV: Reinventing RNNs for the Transformer Era}},
  author = {Peng, Bo and Alcaide, Eric and Anthony, Quentin and Albalak, Alon and Arcadinho, Samuel and Biderman, Stella and Cao, Huanqi and Cheng, Xin and Chung, Michael and Derczynski, Leon and Du, Xingjian and Grella, Matteo and Gv, Kranthi and He, Xuzheng and Hou, Haowen and Kazienko, Przemyslaw and Kocon, Jan and Kong, Jiaming and Koptyra, Bart{\l}omiej and Lau, Hayden and Lin, Jiaju and Mantri, Krishna Sri Ipsit and Mom, Ferdinand and Saito, Atsushi and Song, Guangyu and Tang, Xiangru and Wind, Johan and Wo{\'z}niak, Stanis{\l}aw and Zhang, Zhenyuan and Zhou, Qinghua and Zhu, Jian and Zhu, Rui-Jie},
  year = {2023},
  booktitle = {Findings of the Association for Computational Linguistics: EMNLP 2023},
  publisher = {Association for Computational Linguistics},
  address = {Singapore},
  pages = {14048--14077},
  doi = {10.18653/v1/2023.findings-emnlp.936},
  url = {https://aclanthology.org/2023.findings-emnlp.936/},
  editor = {Bouamor, Houda and Pino, Juan and Bali, Kalika}
}

@misc{qi2026neural,
  title = {{Neural Operator Quantum State: A Foundation Model for Quantum Dynamics}},
  author = {Qi, Zihao and Earls, Christopher and Peng, Yang},
  year = {2026},
  url = {https://arxiv.org/abs/2603.25066},
  eprint = {2603.25066},
  primaryclass = {quant-ph},
  archiveprefix = {arXiv}
}

@article{rende2025foundation,
  title = {{Foundation neural-networks quantum states as a unified Ansatz for multiple hamiltonians}},
  author = {Rende, Riccardo and Viteritti, Luciano Loris and Becca, Federico and Scardicchio, Antonello and Laio, Alessandro and Carleo, Giuseppe},
  year = {2025},
  journal = {Nature Communications},
  volume = {16},
  number = {1},
  pages = {7213},
  doi = {10.1038/s41467-025-62098-x},
  issn = {2041-1723},
  url = {https://doi.org/10.1038/s41467-025-62098-x}
}

@misc{roth2020iterative,
  title = {{Iterative Retraining of Quantum Spin Models Using Recurrent Neural Networks}},
  author = {Roth, Christopher},
  year = {2020},
  url = {https://arxiv.org/abs/2003.06228},
  eprint = {2003.06228},
  archiveprefix = {arXiv},
  primaryclass = {physics.comp-ph}
}

@article{sfedwards1975theory,
  title = {{Theory of spin glasses}},
  author = {{S F Edwards} and {P W Anderson}},
  year = {1975},
  journal = {Journal of Physics F: Metal Physics},
  volume = {5},
  number = {5},
  pages = {965},
  doi = {10.1088/0305-4608/5/5/017},
  issn = {0305-4608},
  url = {https://dx.doi.org/10.1088/0305-4608/5/5/017}
}

@inproceedings{singha2025multilevel,
  title = {{Multilevel Generative Samplers for Investigating Critical Phenomena}},
  author = {Singha, Ankur and Cellini, Elia and Nicoli, Kim Andrea and Jansen, Karl and K{\"u}hn, Stefan and Nakajima, Shinichi},
  year = {2025},
  booktitle = {The Thirteenth International Conference on Learning Representations},
  url = {https://openreview.net/forum?id=YcUV5apdlq}
}

@misc{singha2026scalable,
  title = {{Scalable Generative Sampling and Multilevel Estimation for Lattice Field Theories Near Criticality}},
  author = {Singha, A. and Kauffmann, J. and Cellini, E. and Jansen, K. and Nakajima, S.},
  year = {2026},
  url = {https://arxiv.org/abs/2604.10209},
  eprint = {2604.10209},
  primaryclass = {hep-lat},
  archiveprefix = {arXiv}
}

@misc{sm,
  note = {See Supplemental Material for the derivation of the variational free-energy bound and REINFORCE gradient estimator, details of the encoder-decoder architecture with gauge-fixing and curriculum learning strategy, and extended numerical results for the 2D Edwards-Anderson and random-bond Ising models.}
}

@article{su2024roformer,
  title = {{RoFormer: Enhanced Transformer with Rotary Position Embedding}},
  author = {Su, Jianlin and Ahmed, Murtadha and Lu, Yu and Pan, Shengfeng and Bo, Wen and Liu, Yunfeng},
  year = {2024},
  journal = {Neurocomputing},
  volume = {568},
  pages = {127063},
  doi = {10.1016/j.neucom.2023.127063},
  issn = {0925-2312},
  url = {https://www.sciencedirect.com/science/article/pii/S0925231223011864}
}

@inproceedings{tan2025amortized,
  title = {{Amortized Sampling with Transferable Normalizing Flows}},
  author = {Tan, Charlie B. and Hassan, Majdi and Klein, Leon and Syed, Saifuddin and Beaini, Dominique and Bronstein, Michael M. and Tong, Alexander and Neklyudov, Kirill},
  year = {2025},
  booktitle = {Advances in Neural Information Processing Systems 38: Annual Conference on Neural Information Processing Systems 2025, NeurIPS 2025, San Diego, CA, USA, December 2-7, 2025 / Mexico City, Mexico, November 30 - December 5, 2025},
  url = {http://papers.nips.cc/paper_files/paper/2025/hash/8780dbea239de8fd1401f9ec6721608d-Abstract-Conference.html},
  editor = {Belgrave, Danielle and Zhang, Cheng and Montoya, Laura N. and Lin, Hsuan-Tien and Pascanu, Razvan and Koniusz, Piotr and Ghassemi, Marzyeh and Chen, Nancy and Ru{\'\i}z, Iv{\'a}n Vladimir Meza and {Loaiza-Bonilla}, Arturo},
  bibsource = {dblp computer science bibliography, https://dblp.org}
}

@article{tang2023neuralnetwork,
  title = {{Neural-Network Solutions to Stochastic Reaction Networks}},
  author = {Tang, Ying and Weng, Jiayu and Zhang, Pan},
  year = {2023},
  journal = {Nature Machine Intelligence},
  volume = {5},
  number = {4},
  pages = {376--385},
  doi = {10.1038/s42256-023-00632-6},
  issn = {2522-5839},
  url = {https://doi.org/10.1038/s42256-023-00632-6}
}

@article{tang2024learning,
  title = {{Learning Nonequilibrium Statistical Mechanics and Dynamical Phase Transitions}},
  author = {Tang, Ying and Liu, Jing and Zhang, Jiang and Zhang, Pan},
  year = {2024},
  journal = {Nature Communications},
  volume = {15},
  number = {1},
  pages = {1117},
  doi = {10.1038/s41467-024-45172-8},
  issn = {2041-1723},
  url = {https://doi.org/10.1038/s41467-024-45172-8},
  copyright = {All rights reserved}
}

@article{thouless1977solution,
  title = {{Solution of 'Solvable model of a spin glass'}},
  author = {Thouless, D. J. and Anderson, P. W. and Palmer, R. G.},
  year = {1977},
  journal = {The Philosophical Magazine: A Journal of Theoretical Experimental and Applied Physics},
  publisher = {Taylor \& Francis},
  volume = {35},
  number = {3},
  pages = {593--601},
  doi = {10.1080/14786437708235992},
  issn = {0031-8086},
  url = {https://doi.org/10.1080/14786437708235992}
}

@inproceedings{vaswani2017attention,
  title = {{Attention is All you Need}},
  author = {Vaswani, Ashish and Shazeer, Noam and Parmar, Niki and Uszkoreit, Jakob and Jones, Llion and Gomez, Aidan N and Kaiser, {\L}ukasz and Polosukhin, Illia},
  year = {2017},
  booktitle = {Advances in Neural Information Processing Systems},
  publisher = {Curran Associates, Inc.},
  volume = {30},
  url = {https://proceedings.neurips.cc/paper_files/paper/2017/file/3f5ee243547dee91fbd053c1c4a845aa-Paper.pdf},
  editor = {Guyon, I. and Luxburg, U. Von and Bengio, S. and Wallach, H. and Fergus, R. and Vishwanathan, S. and Garnett, R.}
}

@misc{viteritti2026quantum,
  title = {{Quantum Spin Glass in the Two-Dimensional Disordered Heisenberg Model via Foundation Neural-Network Quantum States}},
  author = {Viteritti, Luciano Loris and Rende, Riccardo and {Bracci-Testasecca}, Giacomo and Niedda, Jacopo and Moessner, Roderich and Carleo, Giuseppe and Scardicchio, Antonello},
  year = {2026},
  url = {https://arxiv.org/abs/2507.05073},
  eprint = {2507.05073},
  primaryclass = {cond-mat.dis-nn},
  archiveprefix = {arXiv}
}

@article{wegner1971duality,
  title = {{Duality in Generalized Ising Models and Phase Transitions without Local Order Parameters}},
  author = {Wegner, Franz J.},
  year = {1971},
  journal = {Journal of Mathematical Physics},
  volume = {12},
  number = {10},
  pages = {2259--2272},
  doi = {10.1063/1.1665530},
  issn = {0022-2488},
  url = {https://doi.org/10.1063/1.1665530},
  urldate = {2026-06-11}
}

@misc{weng2025tracking,
  title = {{Tracking Large Chemical Reaction Networks and Rare Events by Neural Networks}},
  author = {Weng, Jiayu and Zhu, Xinyi and Liu, Jing and L{\"u}, Linyuan and Zhang, Pan and Tang, Ying},
  year = {2025},
  publisher = {arXiv},
  number = {arXiv:2512.10309},
  doi = {10.48550/arXiv.2512.10309},
  url = {http://arxiv.org/abs/2512.10309},
  urldate = {2025-12-16},
  eprint = {2512.10309},
  primaryclass = {q-bio},
  archiveprefix = {arXiv},
  langid = {english}
}

@article{williams1992simple,
  title = {{Simple statistical gradient-following algorithms for connectionist reinforcement learning}},
  author = {Williams, Ronald J.},
  year = {1992},
  journal = {Machine Learning},
  volume = {8},
  number = {3},
  pages = {229--256},
  doi = {10.1007/BF00992696},
  issn = {1573-0565},
  url = {https://doi.org/10.1007/BF00992696}
}

@article{wu2019solving,
  title = {{Solving Statistical Mechanics Using Variational Autoregressive Networks}},
  author = {Wu, Dian and Wang, Lei and Zhang, Pan},
  year = {2019},
  journal = {Physical Review Letters},
  volume = {122},
  number = {8},
  pages = {080602},
  doi = {10.1103/PhysRevLett.122.080602},
  issn = {0031-9007, 1079-7114},
  url = {https://link.aps.org/doi/10.1103/PhysRevLett.122.080602},
  urldate = {2021-03-20},
  langid = {english}
}

@article{wu2021unbiased,
  title = {{Unbiased Monte Carlo cluster updates with autoregressive neural networks}},
  author = {Wu, Dian and Rossi, Riccardo and Carleo, Giuseppe},
  year = {2021},
  journal = {Physical Review Research},
  publisher = {American Physical Society},
  volume = {3},
  number = {4},
  pages = {L042024},
  doi = {10.1103/PhysRevResearch.3.L042024},
  url = {https://link.aps.org/doi/10.1103/PhysRevResearch.3.L042024}
}

@incollection{yedidia2003understanding,
  title = {{Understanding Belief Propagation and Its Generalizations}},
  author = {Yedidia, Jonathan S. and Freeman, William T. and Weiss, Yair},
  year = {2003},
  booktitle = {Exploring Artificial Intelligence in the New Millennium},
  publisher = {Morgan Kaufmann Publishers Inc.},
  address = {San Francisco, CA, USA},
  pages = {239--269},
  isbn = {1-55860-811-7}
}

@article{zhao2025nonequilibrium,
  title = {{Nonequilibrium Statistical Mechanics Revealed by Doob \$h\$ Transform and Variational Autoregressive Networks}},
  author = {Zhao, Yixin and Tang, Ying and Zhang, Pan},
  year = {2025},
  journal = {Physical Review E},
  publisher = {American Physical Society},
  volume = {111},
  number = {3},
  pages = {034120},
  doi = {10.1103/PhysRevE.111.034120},
  url = {https://link.aps.org/doi/10.1103/PhysRevE.111.034120}
}

@misc{zhong2026scalable,
  title = {{Scalable Physics-Inspired Transformers for Spin Glasses}},
  author = {Zhong, Lu and Duan, Wenli and Liu, Jing and Zhang, Pan and Tang, Ying},
  year = {2026},
  url = {https://arxiv.org/abs/2606.22984},
  eprint = {2606.22984},
  archiveprefix = {arXiv},
  primaryclass = {cond-mat.dis-nn}
}

\end{document}


\title{Supplemental Material}

\maketitle
\tableofcontents

\section{Network Architecture and Training Details}
\label{sec:architecture}
\subsection{Variational quenched-averaged free energy}
\label{subsec:variational}
The variational bound $\mathcal{F}(\theta)$ introduced in the main text is derived from the expected Kullback--Leibler (KL) divergence between $q_\theta$ and $P_J$, averaged over the disorder ensemble:
\begin{equation}\label{eq:expected-kl}
    \mathbb{E}_{J \sim P(J)}\left[D_{\mathrm{KL}}(q_\theta(\cdot\vert J) \Vert P(\cdot\vert J))\right] = \mathbb{E}_{J \sim P(J)} \mathbb{E}_{\s\sim q_\theta(\cdot\vert J)} \left[ \ln q_\theta(\s\vert J) + \beta E_J(\s) + \ln Z_J \right].
\end{equation}
Since the partition function $Z_J$ is independent of $\theta$, minimizing the expected KL divergence is equivalent to minimizing $\beta\mathcal{F}(\theta)$, and thus $\mathcal{F}(\theta)$.
Due to the non-negativity of the KL divergence, we have:
\begin{equation}\label{eq:variational-bound}
    \beta \mathcal{F}(\theta) \geq - \mathbb{E}_{J \sim P(J)} \mathbb{E}_{\s\sim q_\theta(\cdot\vert J)} \left[ \ln Z_J \right] = \beta \mathbb{E}_{J \sim P(J)}[F_{\mathrm{true}}(J)],
\end{equation}
where $F_{\mathrm{true}}(J)=-(1/\beta) \ln Z_J$ is the exact free energy of a specific instance $J$. 
Eq.~\eqref{eq:variational-bound} is the variational bound in the main text.
We minimize $\mathcal{F}(\theta)$ by stochastic gradient descent. Since the spin variables are discrete, we employ the score-function (REINFORCE) estimator~\cite{williams1992simple}. For a general expectation over $q_\theta$, we have
\begin{equation}
    \nabla_\theta\,\mathbb{E}_{\s\sim q_\theta}[f(\s)] = \mathbb{E}_{\s\sim q_\theta}\left[ f(\s)\,\nabla_\theta\ln q_\theta(\s) \right].
\end{equation}
Applying this identity to $\mathcal{F}(\theta)$ yields
\begin{equation}\label{eq:sf-grad}
    \nabla_\theta\mathcal{F}(\theta) = \mathbb{E}_{J \sim P(J)}\,\mathbb{E}_{\s\sim q_\theta(\cdot\vert J)} \left[ R_J(\s)\,\nabla_\theta\ln q_\theta(\s\vert J) \right],
\end{equation}
where $R_J(\s) = E_J(\s) + \frac{1}{\beta}\ln q_\theta(\s\vert J)$.
The estimator~\eqref{eq:sf-grad} is unbiased by construction but can exhibit high variance when $R_J(\s)$ fluctuates strongly across samples.
To control this variance, we subtract a disorder-dependent baseline:
\begin{equation}
    \nabla_\theta\mathcal{F}(\theta) = \mathbb{E}_{J \sim P(J)}\,\mathbb{E}_{\s\sim q_\theta(\cdot\vert J)} \left[ \bigl(R_J(\s) - b(J)\bigr)\,\nabla_\theta\ln q_\theta(\s\vert J) \right].
\end{equation}
Subtracting $b(J)$ does not bias the gradient since $\mathbb{E}_{\s\sim q_\theta}[\nabla_\theta\ln q_\theta] = 0$.
In practice, each training step samples $N_{\mathrm{dis}}$ independent disorder realizations $\{J_m\}_{m=1}^{N_{\mathrm{dis}}}$ from $P(J)$. 
Then for each realization, we draw $N_{\mathrm{bs}}$ spin configurations $\{\s_{m,n}\}_{n=1}^{N_{\mathrm{bs}}}$ from $q_\theta(\cdot\vert J_m)$ and compute the reward
\begin{equation}
    R_{m,n} = E_{J_m}(\s_{m,n}) + \frac{1}{\beta}\ln q_\theta(\s_{m,n}\vert J_m).
\end{equation}
The baseline for the $m$-th instance is estimated as $\hat{b}_m = \frac{1}{N_{\mathrm{bs}}}\sum_{n=1}^{N_{\mathrm{bs}}} R_{m,n}$, yielding the gradient estimate
\begin{equation}\label{eq:grad-estimate}
    \nabla_\theta\mathcal{F}(\theta) \approx \frac{1}{N_{\mathrm{dis}}N_{\mathrm{bs}}} \sum_{m=1}^{N_{\mathrm{dis}}}\sum_{n=1}^{N_{\mathrm{bs}}} \left(R_{m,n} - \hat{b}_m\right)\, \nabla_\theta\ln q_\theta(\s_{m,n}\vert J_m).
\end{equation}

\subsection{Encoder}
\label{subsec:encoder}
The encoder maps a disorder realization $J$ to per-site embeddings that capture both the local coupling topology and global frustration patterns.
For a 2D square lattice, the input consists of two coupling tensors $J_h, J_v$ holding the horizontal and vertical bond strengths, respectively.
The encoder exploits the regular lattice geometry through an axial transformer architecture~\cite{ho2019axial}, which alternates row-wise and column-wise self-attention.
This reduces the per-layer cost from $O(N^2) = O(L^4)$ for dense attention to $O(L^3)$, while still providing each site with a global receptive field after two stacked transformer layers.
The encoder is composed of a stack of pre-norm residual blocks.
Each block updates the hidden states by sequentially applying row-wise attention conditioned on the horizontal bonds, column-wise attention conditioned on the vertical bonds, and a standard feed-forward network (FFN). 
Formally, the hidden state $\h_i \in \mathbb{R}^{d_{\mathrm{enc}}}$ is updated as:
\begin{equation}\label{eq:encoder-block}
\begin{aligned}
    \h_i &\leftarrow \h_i + \operatorname{RowAttn}(\operatorname{LN}(\h_i); J_h), \\
    \h_i &\leftarrow \h_i + \operatorname{ColAttn}(\operatorname{LN}(\h_i); J_v), \\
    \h_i &\leftarrow \h_i + \operatorname{FFN}(\operatorname{LN}(\h_i)),
\end{aligned}
\end{equation}
where $\operatorname{LN}$ denotes layer normalization. 
After passing through all encoder blocks and a final layer normalization, the network outputs the continuous site embeddings $\mathbf{H} = [\mathbf{h}_1, \dots, \mathbf{h}_N] \in \mathbb{R}^{N \times d_{\mathrm{enc}}}$, which subsequently serve as the keys and values for the decoder's cross-attention mechanism.
Within the axial transformer blocks, coupling information is injected into each attention operation through two complementary pathways.
First, an MLP denoted as $\Phi$ computes an additive bias from the scalar couplings, which is applied directly to the pre-softmax attention logits between adjacent sites. 
Second, another MLP, $\Psi$, maps the couplings to continuous feature vectors that are added to the corresponding value representations at the bond's endpoints. 
Consequently, the modified attention operation for a single head takes the general form $\operatorname{Attn}(Q, K, V) = \operatorname{softmax}\left(\frac{Q K^\top}{\sqrt{d_{\mathrm{enc}}}} + B_{\Phi}\right) V_{\Psi}$,
where $B_{\Phi}$ is the bias matrix assembled via $\Phi$, and $V_{\Psi}$ represents the value matrix enriched by the additive bond features from $\Psi$. 

\begin{figure}[!t]
\centering
\includegraphics[width=0.5\linewidth]{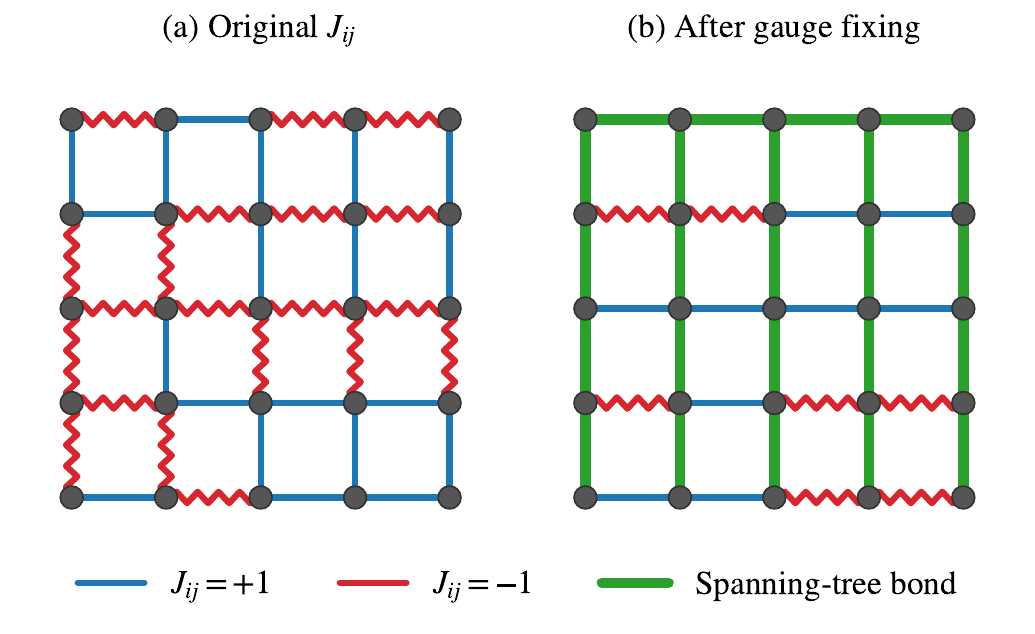}
\caption{Gauge-fixing via spanning-tree canonicalization.
Left: Raw coupling matrix $\{J_{ij}\}$ drawn from the bimodal distribution (blue: ferromagnetic $J_{ij}=+1$, red: antiferromagnetic $J_{ij}=-1$).
Right: Gauge-fixed couplings after applying $J_{ij} \leftarrow \epsilon_i \epsilon_j J_{ij}$ with signs propagated along the breadth-first spanning tree (tree edges highlighted in green). All tree bonds become ferromagnetic, while non-tree bonds encode the residual physical frustration.}
\label{fig:gauge}
\end{figure}

\begin{figure}[!t]
\centering
\includegraphics[width=0.7\linewidth]{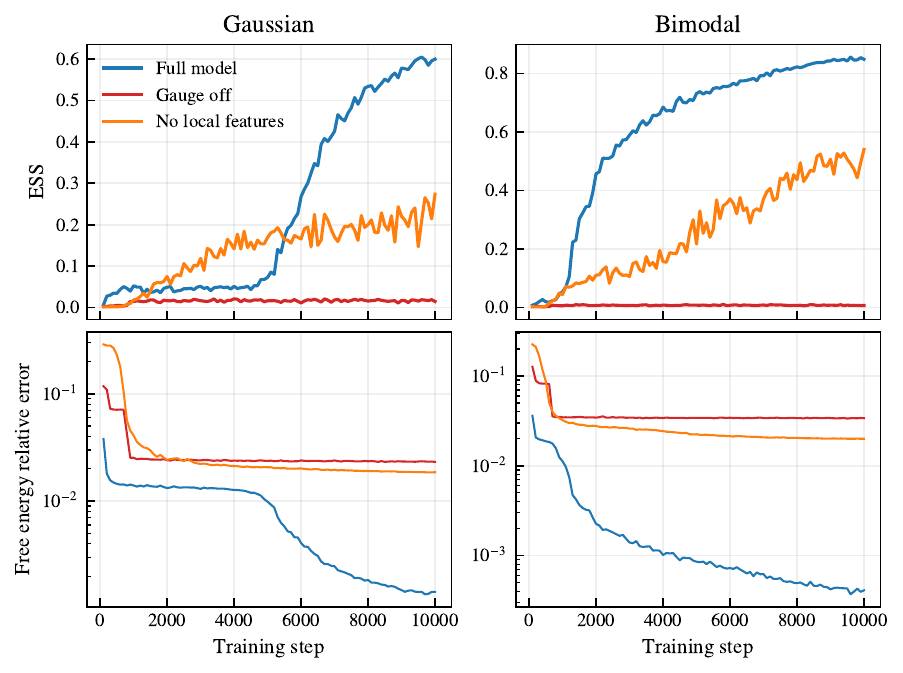}
\caption{Ablation experiments on training dynamics for $L=12$ with $\beta=1.0$.
Top: Effective sample size (ESS) as a function of training steps for Gaussian (left) and bimodal (right) disorder.
Bottom: Relative error of the variational free energy for the same two disorder ensembles.
Three model variants are compared: the full model with gauge-fixing and the local coupling bypass (Full model), the model without gauge-fixing (Gauge off), and the model with the local coupling bypass removed (No local features).
Removing either component degrades both sample quality and free-energy accuracy.}
\label{fig:ablation}
\end{figure}

The encoder must operate on a canonical gauge-fixed coupling matrix.
We fix the gauge by building a breadth-first spanning tree on the $L \times L$ lattice rooted at site $0$, which connects all $N$ sites without forming any cycles.
Gauge signs are initialized as $\epsilon_0 = +1$ at the root and propagated outward: for each tree edge connecting a parent site $i$ to its child $j$, the child sign is set to $\epsilon_j = \epsilon_i \cdot \operatorname{sgn}(J_{ij})$, so that after the canonical transformation $J_{ij} \leftarrow \epsilon_i \epsilon_j J_{ij}$, every tree bond becomes ferromagnetic.
The same transformation is applied to all non-tree bonds. Because these bonds close cycles in the lattice, they are not forced to a particular sign and instead encode the genuine physical frustration.
Because the tree is cycle-free, this assignment is always consistent and collapses the $2^N$-fold degeneracy to a single representative, up to a global flip of all $\epsilon_i$.
Figure~\ref{fig:gauge} illustrates the procedure: the raw coupling matrix (left) is mapped to its gauge-fixed counterpart (right), with spanning tree bonds drawn in green.
Without this canonicalization procedure, the network wastes representational capacity navigating the $2^N$-fold $\mathbb{Z}_2$ gauge degeneracy. 
As shown in Fig.~\ref{fig:ablation}, omitting gauge-fixing prevents the training from converging.

\subsection{Decoder}
\label{subsec:decoder}
The decoder is a causal transformer that autoregressively generates spin configurations conditioned on the global disorder embeddings $\mathbf{H}$. 
To reduce the sequence length, the physical spins $\mathbf{s}$ are grouped into coarse-grained $p_x \times p_y$ patches, represented by a sequence of categorical variables $\x = (x_1, \dots, x_{N_{\mathrm{patch}}})$ where $N_{\mathrm{patch}} = L^2 / (p_x \times p_y)$.
The generation follows a patch-level autoregressive factorization $q_\theta(\s | J) = \prod_{t=1}^{N_{\mathrm{patch}}} q_\theta(x_t | x_{<t}, J)$. 
Building upon the architecture of FlashVAN~\cite{zhong2026scalable}, the decoder leverages FlashAttention and a physics-informed sparse attention mechanism.
However, in contrast to the decoder-only FlashVAN, our architecture introduces a cross-attention mechanism to explicitly condition the autoregressive generation on the global disorder embeddings $\mathbf{H}$.
At each step $t$, the hidden state $\g_t \in \mathbb{R}^{d_{\mathrm{emb}}}$ is updated via causally masked self-attention, followed by cross-attention with $\mathbf{H}$, and an FFN:
\begin{equation}
\begin{aligned}
    \g_t &\leftarrow \g_t + \operatorname{SelfAttn}(\operatorname{LN}(\g_t)), \\
    \g_t &\leftarrow \g_t + \operatorname{CrossAttn}(\operatorname{LN}(\g_t), \mathbf{H}), \\
    \g_t &\leftarrow \g_t + \operatorname{FFN}(\operatorname{LN}(\g_t)).
\end{aligned}
\end{equation}
At each autoregressive step $t$, the decoder input $\g_t \in \mathbb{R}^{d_{\mathrm{emb}}}$ is assembled by concatenating two embeddings $\operatorname{Concat}(\mathbf{e}_s^{(t)}, \mathbf{e}_J^{(t)})$: a learned spin token embedding $\mathbf{e}_s^{(t)} = \mathrm{Embed}(x_{t-1})$ that maps the discrete patch configuration from the preceding step (a zero start token for $t=0$) to a continuous vector, and a local coupling embedding $\mathbf{e}_J^{(t)}$ obtained by linearly projecting the incident bond couplings (left, right, up, down) of the sites within the $t$-th patch.
This local coupling bypass is conceptually redundant: cross-attention over the global disorder embeddings $\mathbf{H}$ can in principle retrieve any local bond quantity the decoder requires. 
Empirically, however, removing this bypass substantially increases training difficulty and degrades generalization to unseen lattice sizes (Fig.~\ref{fig:ablation}).
We conjecture that the explicit local features serve a role analogous to a residual connection, providing a direct gradient pathway for local coupling information.

Here, the decoder employs rotary position embeddings (RoPE)~\cite{su2024roformer} rather than the absolute or learnable positional encodings adopted in FlashVAN~\cite{zhong2026scalable}.
RoPE encodes relative patch positions through rotation matrices applied to the query and key vectors within self-attention.
This confers translation invariance on the patch grid and naturally supports zero-shot transfer across lattice sizes, since relative-position encoding does not depend on absolute grid coordinates.
Finally, a linear projection maps the processed decoder state to logits over the $2^{p_x p_y}$ possible patch configurations: $q_\theta(x_t \mid x_{<t}, J) = \operatorname{softmax}(W_{\mathrm{out}} \g_t)$.

\subsection{Curriculum learning strategy}
\label{subsec:curriculum}
\begin{figure}[!ht]
\centering
\includegraphics[width=0.7\linewidth]{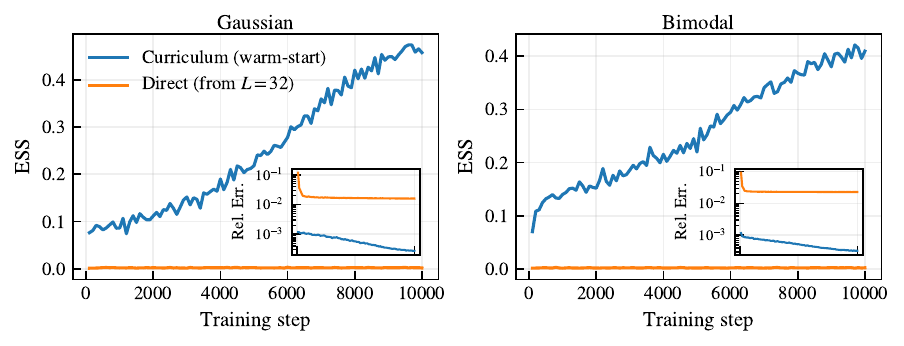}
\caption{Curriculum learning versus direct optimization at $L=32$ for bimodal disorder ($\beta=1.0$). The curriculum run (blue) is warm-started from a converged model at $L=24$, while the direct run (red) is trained from random initialization. The direct run fails to escape the initial plateau, with the ESS remaining near zero, whereas the curriculum run rapidly converges to a high-quality solution.}
\label{fig:curriculum-example}
\end{figure}

Optimizing the variational free energy for large-scale spin glasses from random initialization is notoriously difficult~\cite{zhong2026scalable,bialas2026sampling}.
As demonstrated in the main text (Fig.~2c), our size-agnostic encoder-decoder architecture already exhibits non-trivial zero-shot transfer: a model trained on $L=12$, when evaluated directly on larger lattices without retraining, outperforms both the naive mean-field baseline and a randomly initialized model.
This indicates that the learned representations capture transferable correlations across length scales, and that a model pre-trained at a smaller size provides a markedly better initialization than random weights for larger systems.
We exploit this property through a curriculum learning strategy~\cite{bengio2009curriculum}: the model is trained sequentially across increasing system sizes $\{8,12,16,24,32,40,48\}$, with each stage warm-started from the converged weights of the previous stage.
As shown in Fig.~\ref{fig:curriculum-example}, for $L=32$ at $\beta=1.0$, direct optimization from a randomly initialized model fails to learn, with the ESS remaining near zero, whereas the curriculum run warm-started from $L=24$ converges to a high-fidelity solution.

\subsection{Training details}
\label{subsec:training}
The same model architecture is used for both bimodal and Gaussian disorder distributions, and Table~\ref{tab:hyperparameters} summarizes the key hyperparameters. 
The total number of trainable parameters is approximately \num{0.9}\,M.
Training minimizes the variational quenched-averaged free energy using the REINFORCE gradient estimator with a per-disorder baseline, as derived in Sec.~\ref{subsec:variational}.
At each training step, $N_{\mathrm{dis}} = 20$ independent disorder realizations are drawn from $P(J)$, and for each realization, $N_{\mathrm{bs}} = 256$ spin configurations are sampled to compute the gradient estimate in Eq.~\eqref{eq:grad-estimate}.
The encoder and decoder are jointly optimized using the Muon optimizer for hidden-weight matrices and AdamW for all remaining parameters, with learning rates $2\times 10^{-2}$ and $10^{-3}$, respectively.
A cosine schedule decays the learning rates to $0.1$ of their initial values over $10^4$ training steps per system size, and gradients are clipped to a maximum norm of $1$.
Gauge fixing (Sec.~\ref{subsec:encoder}) is applied to every disorder realization before encoding.
For a fixed $L$, training for $10^4$ steps takes approximately \qty{0.45}{h} at $L=8$ and \qty{21}{h} at $L=48$ on a single RTX 5090 GPU.
The curriculum warm-start strategy that chains multiple system sizes is described in Sec.~\ref{subsec:curriculum}.

For ground-state search, we employ a variational annealing schedule~\cite{hibat-allah2021variational} that gradually cools the model from the finite-temperature regime to the zero-temperature limit.
The model is initialized from a converged finite-temperature checkpoint, and the temperature is then linearly decreased over $N_{\mathrm{anneal}} = 100$ annealing steps as $T(k) = T_0\,(1 - k / N_{\mathrm{anneal}})$, with $k = 0, \dots, N_{\mathrm{anneal}}-1$.
At each temperature step, the network is trained for $N_{\mathrm{train}} = 100$ gradient steps, yielding $10^4$ total steps.
As $T \to 0$, the entropy term is suppressed and the variational objective reduces to pure energy minimization, forcing the network to concentrate its probability mass on the lowest-energy configurations.
All other training hyperparameters are identical to those used for finite-temperature training.
After annealing, ground-state candidates are obtained by drawing \num{1024} independent samples per disorder realization and retaining the configuration with the lowest energy.
Exact ground-state energies for validation are computed via the method of~\cite{pardella2008exact}.
The same procedure is applied to both bimodal and Gaussian disorder.

\begin{table}[!ht]
\centering
\caption{Model architecture hyperparameters. The same configuration is used for both bimodal and Gaussian disorder.}
\label{tab:hyperparameters}
\begin{tabular}{llc}
\toprule
Notation & Hyperparameter & Value \\
\midrule
\multicolumn{3}{c}{Encoder} \\
\midrule
$d_{\mathrm{enc}}$ & Hidden dimension & 64 \\
$N_{\mathrm{enc}}$ & Number of axial transformer blocks & 2 \\
$N_{\mathrm{head}}^{\mathrm{enc}}$ & Number of attention heads & 4 \\
$d_{\mathrm{MLP}}^{\mathrm{enc}}$ & MLP hidden dimension & 256 \\
\midrule
\multicolumn{3}{c}{Decoder} \\
\midrule
$d_{\mathrm{emb}}$ & Embedding dimension & 128 \\
$N_{\mathrm{dec}}$ & Number of decoder blocks & 3 \\
$N_{\mathrm{head}}^{\mathrm{dec}}$ & Number of attention heads & 8 \\
$d_{\mathrm{MLP}}^{\mathrm{dec}}$ & MLP hidden dimension & 512 \\
$d_{\mathrm{token}}$ & Spin token embedding dimension & 64 \\
$d_{\mathrm{local}}$ & Local coupling feature dimension & 64 \\
$(p_x, p_y)$ & Patch size & $(2, 2)$ \\
\bottomrule
\end{tabular}
\end{table}

\section{Extended Numerical Results}
\label{sec:results}
\subsection{2D EA model}
\begin{figure}[!ht]
\centering
\includegraphics[width=0.7\linewidth]{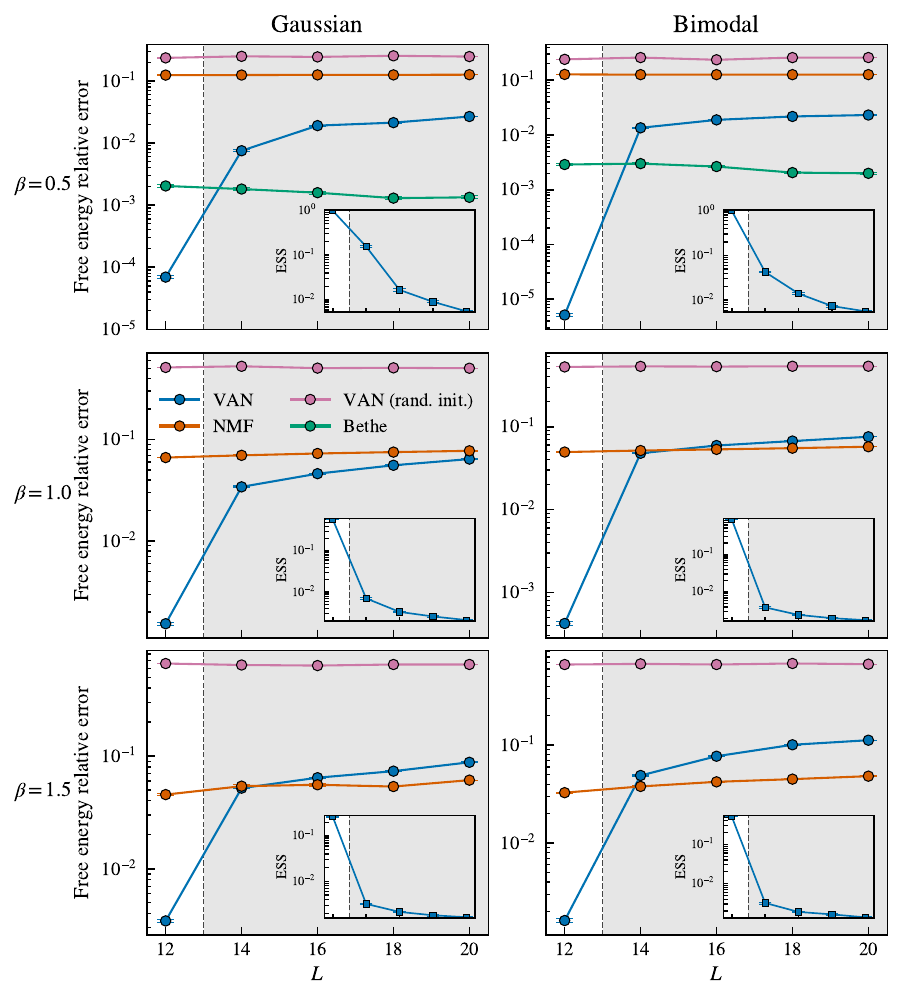}
\caption{System-size extrapolation across disorder types and temperatures.
The model is trained on $L=12$ lattices and evaluated on sizes $L = 12$-$20$ without retraining (grey shading marks the extrapolation regime $L>12$).
Rows correspond to inverse temperatures $\beta = 0.5, 1.0, 1.5$; columns correspond to Gaussian (left) and bimodal (right) disorder distributions.
For $\beta=0.5$, the Bethe approximation provides an additional reference.
At $\beta=1.0$ and $1.5$, belief propagation fails to converge on a significant fraction of instances.
The Gaussian, $\beta=1.0$ panel (center left) reproduces the data shown in Fig.~2(c) of the main text.}
\label{fig:size-gen}
\end{figure}

Here we provide extended numerical results for the 2D EA model.
Figure~\ref{fig:size-gen} provides a comprehensive extension of the size-generalization analysis presented in Fig.~2(c) of the main text. 
While the main text shows only the Gaussian-disorder case at $\beta=1.0$, here we evaluate the model across both Gaussian and bimodal couplings at three representative inverse temperatures ($\beta = 0.5, 1.0, 1.5$). 
The model is trained exclusively on $L=12$ lattices and directly applied, without fine-tuning, to larger systems up to $L=20$, with grey shading indicating the extrapolation regime $L>12$.
At $\beta=0.5$, we additionally benchmark against the Bethe free energy computed via belief propagation~\cite{yedidia2003understanding}.
This comparison is omitted for $\beta=1.0$ and $1.5$ because belief propagation fails to converge on a significant fraction of the test instances, due to the presence of loops in the 2D lattice and the stronger frustration at lower temperatures.
At small to moderate $L$, the trained VAN outperforms both the naive mean-field (NMF) approximation and a randomly initialized VAN in free-energy accuracy and sampling efficiency. 
As $L$ increases, this advantage progressively diminishes: the relative error approaches the NMF baseline and the ESS drops by over an order of magnitude.
In most cases, however, the VAN remains competitive with or better than NMF.
Two systematic trends are evident across all panels.
First, the degradation is monotonic in $L$. 
Second, it is more pronounced at lower temperatures ($\beta=1.5$), where the increasingly rugged free-energy landscape impedes the transfer of long-range correlations.
We also emphasize that the low ESS values at larger $L$ in Fig.~\ref{fig:size-gen} reflect direct zero-shot evaluation without retraining. 
As demonstrated in Fig.~\ref{fig:curriculum-example}, fine-tuning the pre-trained model at the target system size rapidly restores high ESS values, confirming that the zero-shot representations serve as an effective warm-start. 

\begin{figure}[!ht]
\centering
\includegraphics[width=0.7\linewidth]{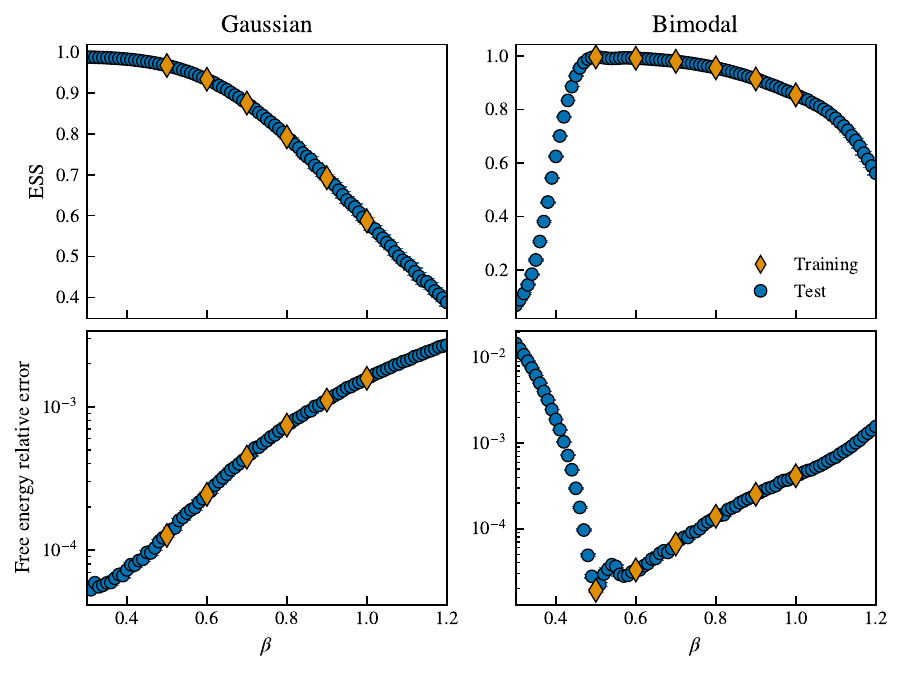}
\caption{Thermal generalization across disorder types.
The encoder is conditioned on $K_{ij} = \beta J_{ij}$ and trained on a discrete set of $\beta$ values.
For Gaussian couplings the ESS decays monotonically with $\beta$, while for bimodal couplings the ESS degrades at both temperature extremes (see text). 
The Gaussian ESS panel (top left) reproduces the data shown in Fig.~2(b) of the main text.}
\label{fig:thermal-gen}
\end{figure}

Figure~\ref{fig:thermal-gen} extends the thermal generalization analysis of Fig.~2(b) in the main text.
Rather than training at a fixed temperature, at each training step both a disorder realization $J$ and an inverse temperature $\beta$ are drawn, so the encoder receives the dimensionless coupling $K_{ij} = \beta J_{ij}$ as input. 
The model is trained on a discrete set of $\beta$ values and evaluated across a continuous range of $\beta$. 
While Fig.~2(b) reports only the ESS for Gaussian disorder, here we present the complete picture: ESS (top row) and relative error of the variational free energy (bottom row), for both Gaussian (left column) and bimodal (right column) disorder distributions.
For Gaussian disorder, the ESS decreases monotonically with $\beta$ and the relative error correspondingly increases, consistent with the expectation that lower temperatures present harder sampling problems. 
For bimodal disorder, a qualitatively different behavior emerges: the ESS degrades at both temperature extremes, dropping sharply in the high-temperature extrapolation regime ($\beta \lesssim 0.5$) as well as at low temperatures.
This high-temperature drop is counterintuitive, and we conjecture that the degradation arises because the dimensionless couplings satisfy $|K_{ij}| = \beta$ for discrete $\pm 1$ bonds. 
When extrapolating to $\beta$ values below the training range, all $|K_{ij}|$ shrink uniformly below any magnitude the encoder has encountered, thus hindering its ability to resolve differences between disorder realizations. 
For Gaussian disorder, by contrast, the continuous distribution of bare $J_{ij}$ naturally exposes the encoder to small $K_{ij}$ values already at the training $\beta$.

\begin{figure}[!ht]
\centering
\includegraphics[width=0.7\linewidth]{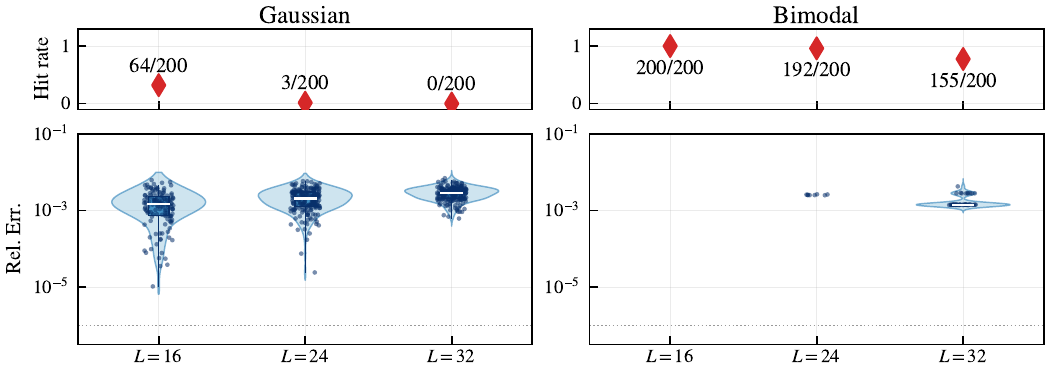}
\caption{Ground-state search performance on Gaussian vs.\ bimodal 2D EA instances (200 instances per size $L$), obtained via the variational annealing protocol described in Sec.~\ref{subsec:training}.
(Top) Hit rate for successfully sampling the exact ground state, with exact energies computed via Ref.~\cite{pardella2008exact}.
(Bottom) Relative error $\Delta E/E_0$ distributions shown as violin plots.
While the network achieves near-perfect exact sampling for discrete bimodal couplings across all sizes, continuous Gaussian couplings present a harder generalization task: the hit rate decays and the error distribution broadens systematically as $L$ increases.}
\label{fig:gs-violin}
\end{figure}

While the main text demonstrates the efficiency of our annealed encoder-decoder framework for ground-state sampling on bimodal EA spin glasses up to $L=32$ (Fig.~3(b)), it is important to assess the model's performance on continuous disorder distributions.
Figure~\ref{fig:gs-violin} provides a comparative analysis of the ground-state search capabilities for both Gaussian and bimodal couplings at system sizes $L=16, 24, 32$.
As illustrated, finding the exact ground state for Gaussian instances is fundamentally more challenging.
The network achieves a near-perfect hit rate for bimodal instances across all tested sizes, yet its success rate on Gaussian instances decays systematically as $L$ increases.
We attribute this performance gap to the representational demands on the encoder.
For bimodal instances, the coupling variables $\{J_{ij}\}$ are strictly binary, and the encoder operates on a discrete, finite input space from which it can robustly learn generalizable representations of the local frustration landscape.
In contrast, Gaussian couplings are drawn from a continuous distribution, creating an infinitely large input space.
Identifying the ground state across this continuous landscape requires substantially higher representational capacity.
While the framework still identifies highly competitive low-energy states for Gaussian disorder (with relative errors $\Delta E/E_0 \sim 10^{-3}$), hitting the absolute ground state is substantially harder than in the bimodal case.
A practical remedy is to post-process the model's low-energy candidates with a rapid local-search heuristic, or to use the amortized model as a warm-start for brief per-instance variational annealing~\cite{hibat-allah2021variational}.

\subsection{2D random-bond Ising model}
For the random-bond Ising model, simulations are performed on the Nishimori line~\cite{nishimori1981internal}, where the inverse temperature $\beta$ is fixed by the antiferromagnetic bond probability $p$ via $e^{-2\beta} = p / (1-p)$.

In Table~\ref{tab:binder}, we provide the numerical values of the Binder cumulant $U$ in the main text.
To obtain small statistical errors, we generate $N_{\mathrm{ins}} = 10^6$ distinct disorder realizations for each bond probability $p$ and system size $L$. 
This massive sampling is made computationally feasible by our amortized architecture, which enables zero-shot generation across novel instances after the model is trained.
Our approach thereby circumvents both the prohibitive instance-specific retraining required by standard VANs and the severe critical slowing down and long equilibration times that plague traditional MCMC and parallel tempering algorithms in this regime.
For each instance, we draw independent configurations from our trained model to evaluate the thermal expectation values via neural importance sampling~\cite{nicoli2020asymptotically}:
\begin{equation}
    \langle \mathcal{O} \rangle \approx \frac{\langle w(\s) \mathcal{O}(\s) \rangle_{q_\theta}}{\langle w(\s) \rangle_{q_\theta}},
\end{equation}
where $w(\s) = e^{-\beta E(\s)} / q_\theta(\s)$ is the unnormalized importance weight, and $\mathcal{O} \in \{m^2, m^4\}$.
After computing these instance-specific thermal averages, the quenched averages $[\langle m^2 \rangle]$ and $[\langle m^4 \rangle]$ are obtained by taking the arithmetic mean over the $N_{\mathrm{ins}}$ instances. The Binder cumulant is then evaluated as $U = [\langle m^4 \rangle] / [\langle m^2 \rangle]^2$, with its statistical uncertainty estimated via the bootstrap method over the disorder realizations.

\begin{table}[!ht]
\centering
\caption{Values of Binder cumulant $U$ for different system sizes $L$ and parameters $p$.}
\begin{tabular}{lcccc}
\toprule\toprule
\multirow{2}{*}{$p$} & \multicolumn{4}{c}{$U$} \\ \cmidrule{2-5}
       & $L=6$       & $L=8$       & $L=12$      & $L=16$      \\
\midrule
0.1080 & 1.12096(20) & 1.11999(19) & 1.11830(19) & 1.11721(19) \\
0.1085 & 1.12308(20) & 1.12254(20) & 1.12166(20) & 1.12109(20) \\
0.1090 & 1.12521(20) & 1.12513(20) & 1.12488(20) & 1.12515(20) \\
0.1095 & 1.12740(20) & 1.12770(20) & 1.12811(21) & 1.12906(20) \\
0.1100 & 1.12962(21) & 1.13039(21) & 1.13175(21) & 1.13327(21) \\
0.1105 & 1.13188(21) & 1.13300(21) & 1.13514(21) & 1.13751(21) \\
\bottomrule\bottomrule
\end{tabular}
\label{tab:binder}
\end{table}

\subsection{Computational cost}
We analyze the computational cost of our encoder-decoder architecture, following the FLOPs decomposition of Ref.~\cite{bialas2026sampling}.
For a single forward pass through a decoder-only transformer, the leading-order FLOPs is given by~\cite{bialas2026sampling}
\begin{equation}
    O\!\left(d_{\mathrm{emb}} \cdot 2^{p_x p_y} \cdot \frac{L^2}{p_x p_y}\right)
    + O\!\left(d_{\mathrm{emb}}^2 \cdot \frac{L^2}{p_x p_y}\right)
    + O\!\left(\frac{L^4}{(p_x p_y)^2} \cdot d_{\mathrm{emb}}\right).
    \label{eq:flops-decoder}
\end{equation}
The three terms reflect the final linear projection, the FFN in each decoder block, and the self-attention mechanism.
Note that for OBC, where a physics-informed sparse attention is adopted~\cite{zhong2026scalable}, the third term reduces to $O\!\left(\frac{L^3}{(p_x^2 p_y)} \cdot d_{\mathrm{emb}}\right)$.

Our encoder-decoder architecture introduces two additional components beyond the decoder-only baseline: 
\begin{enumerate}
    \item An axial-transformer encoder.
    The encoder processes the $L \times L$ lattice with row- and column-wise attention, requiring $O(L^3 \cdot d_{\mathrm{enc}})$ operations per forward pass, which is subdominant to the $O(L^4)$ decoder terms for large $L$.
    \item The cross-attention between the $N_{\mathrm{patch}}=L^2/(p_x p_y)$ decoder tokens and the $L^2$ encoder output tokens. It contributes $O(N_{\mathrm{patch}} \cdot L^2 \cdot d_{\mathrm{emb}}) = O(L^4 / (p_x p_y) \cdot d_{\mathrm{emb}})$ operations, which shares the same $O(L^4)$ scaling as the self-attention term but enters with a different power of the patch size.
\end{enumerate}
Since the cross-attention term is always $O(L^4)$, the three-term FLOPs structure of Eq.~\eqref{eq:flops-decoder} holds for both OBC and PBC in our encoder-decoder architecture.
Furthermore, the FLOPs for ancestral sampling with a key-value (KV) cache is the same as Eq.~\eqref{eq:flops-decoder}.
Without a KV cache, the sampling cost would grow as $O(L^6)$, which is prohibitively expensive for large $L$.
All experiments in this work use a KV cache.

While the FLOPs analysis captures the asymptotic behavior, the actual wall time crosses over between two regimes.
When $N_{\mathrm{patch}}$ is small, the FFN dominates the runtime, giving $L^2$ scaling.
When $N_{\mathrm{patch}}$ exceeds $\sim 4 d_{\mathrm{emb}}$, however, the $O(N_{\mathrm{patch}}^2)$ attention operations become the bottleneck in both FLOPs and wall time, and the scaling crosses over to $L^4$.
With our model hyperparameters ($d_{\mathrm{emb}} = 128$, $p_x = p_y = 2$, see Table~\ref{tab:hyperparameters}), this crossover occurs near $L \approx 45$.

\begin{figure}[!ht]
\centering
\includegraphics[width=0.7\linewidth]{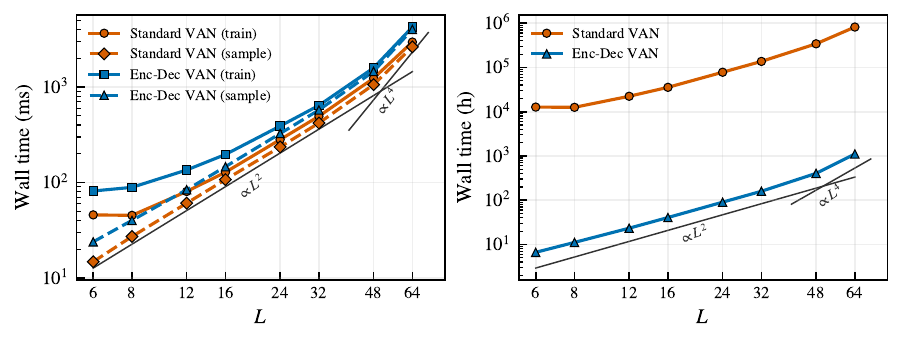}
\caption{
(a) Measured wall time for the standard decoder-only VAN (as in Ref.~\cite{zhong2026scalable}) and our encoder-decoder VAN.
We present per-epoch training time and per-instance sampling time as functions of $L$, with $\propto L^2$ and $\propto L^4$ reference lines.
The time is measured on an RTX 5090 GPU, with the model architecture hyperparameters listed in Table~\ref{tab:hyperparameters} and a batch size of $N_{\mathrm{bs}}=1024$.
(b) Estimated total wall time for $10^6$ disorder realizations.
For the standard VAN we assume $10^3$ training epochs per instance for illustration.}
\label{fig:runtime-rbim}
\end{figure}

Figure~\ref{fig:runtime-rbim} presents the measured runtime data.
Figure~\ref{fig:runtime-rbim}(a) shows the per-epoch training time and per-instance sampling time for both architectures as functions of $L$, tested on the 2D random-bond Ising model with PBC.
At small $L$, all curves follow $\propto L^2$ scaling, consistent with FFN-dominated execution.
From $L=48$, the curves begin to bend upward from the $L^2$ trend line, consistent with the expected crossover toward $L^4$ attention-dominated scaling.
Fig.~\ref{fig:runtime-rbim}(b) translates these per-instance costs into an estimated total wall time for the RBIM study.
For each $(p, L)$ combination, we evaluate $N_{\mathrm{ins}} = 10^6$ independent disorder realizations.
In our encoder-decoder VAN, training is performed once on an ensemble of disorder realizations. The total cost is therefore dominated by the per-instance sampling time multiplied by $N_{\mathrm{ins}}$.
For the standard VAN, each new instance requires training from scratch, and the total cost scales as the training time per epoch multiplied by $N_{\mathrm{ins}} \times N_{\mathrm{epochs}}$.
We set $N_{\mathrm{epochs}} = 10^3$ solely to illustrate the qualitative reduction in computational cost, since the actual epochs required for a converged standard VAN would depend on $L$ and the difficulty of the instance.
The crucial distinction is the functional form of the cost: the encoder-decoder framework replaces per-instance training with amortized inference, reducing the dependence on the number of instances from $O(N_{\mathrm{ins}} \cdot N_{\mathrm{epochs}})$ to $O(N_{\mathrm{ins}})$.
For $N_{\mathrm{ins}} = 10^6$ and any reasonable $N_{\mathrm{epochs}}$, this renders the study feasible, whereas the per-instance training cost of the standard VAN would be prohibitive.

\bibliography{references}